\documentclass[10pt]{article}
\usepackage[margin=0.8in]{geometry}
\usepackage{authblk}
\usepackage{sectsty}
\usepackage{amsmath}
\usepackage{graphicx}
\usepackage{enumerate}
\usepackage{natbib}
\usepackage{url}
\usepackage{hyperref}
\usepackage{amsfonts,amssymb}
\usepackage{subfig,etoolbox}
\sectionfont{\large}
\subsectionfont{\large}
\numberwithin{equation}{section}

\numberwithin{def1}{section}

\numberwithin{res1}{section}

\numberwithin{prop1}{section}

\title{\fontsize{15}{18}\selectfont \textbf{Factor copula models for non-Gaussian longitudinal data}}
\author[]{\fontsize{10}{12}\selectfont Subhajit Chattopadhyay}
\affil[]{\small Bhubandanga, Bidya Sagar Path, Bolpur, West Bengal, India, PIN: 731204}
\date{}
\begin{document}
\bigskip
\maketitle
\begin{abstract}
This article introduces factor copula approaches for modeling the temporal dependence of non-Gaussian (continuous/discrete) longitudinal data. Factor copula models, which are canonical vine copulas, elucidate the underlying dependence structure of multivariate data through latent variables, thereby facilitating interpretation and implementation for unbalanced longitudinal data. We develop regression models for continuous, binary, and ordinal longitudinal data, incorporating covariates, using factor copula constructions with subject-specific latent variables. With consideration for homogeneous within-subject dependence, our proposed models enable feasible parametric inference in moderate to high dimensional scenarios, employing a two-stage (IFM) estimation method. We evaluate the finite sample performance of the proposed models through extensive simulation studies. In empirical analysis, we apply the proposed models to analyze various longitudinal responses from two real-world datasets. Furthermore, we compare the performance of these models with widely used random effect models using standard model selection techniques and observe substantial improvements. Our findings suggest that factor copula models can serve as viable alternatives to random effect models, offering deeper insights into the temporal dependence of longitudinal data across diverse contexts.
\end{abstract}
\section{Introduction}
Longitudinal studies involve the systematic collection of repeated measurements from subjects over time, aimed at understanding their relationship with explanatory variables (covariates) and uncovering the underlying mechanisms of dependence among these measurements. Depending on the research focus, these measurements can manifest in various forms, ranging from discrete to continuous. While statistical methodologies for continuous longitudinal data have been extensively explored, benefiting from a rich array of flexible multivariate distributions, approaches for handling discrete cases remain relatively scarce. Seminal works by researchers such as \citet{verbeke1997linear}, \citet{davis2002statistical}, and \citet{fitzmaurice2008longitudinal} offer invaluable insights and foundational concepts in longitudinal data analysis. Linear mixed models (LMM) have emerged as the go-to choice for analyzing continuous longitudinal data, resting on the assumption of multivariate normality for the repeated measurements (\citet{laird1982random}). However, the landscape has seen the emergence of alternative methodologies aiming to relax stringent distributional assumptions and accommodate non-Gaussian longitudinal data. Among these, the copula framework has garnered increasing attention in recent years for its versatility and ability to capture complex dependencies within longitudinal data.
\par The application of copulas to longitudinal data analysis traces back to \citet{lambert2002copula}, marking an early utilization of this methodology in this context. \citet{sun2008heavy} proposed the utilization of multivariate elliptical copulas to model longitudinal data, contributing to the diversification of copula applications. \citet{shi2016multilevel} delved into unbalanced longitudinal study designs, employing Gaussian copulas to address the intricacies of such setups. More recently, \citet{killiches2018ad} adopted a flexible D-vine copula approach to model unbalanced continuous longitudinal data, leveraging linear mixed models as marginals. While D-vine copulas offer remarkable flexibility and naturally account for the temporal ordering of repeated measurements, they can pose computational challenges in high-dimensional settings due to the involvement of numerous parameters. \citet{panagiotelis2012pair} explored the use of pair-copulas to model dependence in discrete data, devoid of covariate involvement, whereas \citet{zilko2016copula} employed pair-copulas to address mixed data scenarios. \citet{sefidi2021pair} extended the application of D-vine copulas to construct various multivariate distributions using power series marginal distributions, albeit within a limited dimensionality ($d = 3$). Factor copulas, introduced by \citet{krupskii2013factor}, offer comparable flexibility with fewer dependence parameters, explaining the interrelationships between observed variables through latent variables. \citet{nikoloulopoulos2015factor} extended the factor copula methodology to model discrete-valued item response data, while \citet{kadhem2021factor} applied factor copula models to mixed response-type social science data, further broadening the scope of applications in this field.
\par Factor copula models offer distinct advantages, particularly their applicability to response variables of any nature without encountering numerical complexities, even in moderate to high dimensions. These models, classified as truncated vine-copula models, incorporate both observed and latent variables. By selecting appropriate bivariate linking copulas, factor copulas afford a broad spectrum of asymmetric, tail, and nonlinear dependence structures to the data. \citet{kadhem2021factor} underscored the interpretability and superior fit of factor copula models compared to vine copula models, noting their closure under marginals. This property ensures that lower-order marginals belong to the same parametric copula family, and different permutations of observed variables yield identical distributions. In this article, our primary contribution lies in introducing factor copulas as a tool for modeling the temporal dependence inherent in unbalanced non-Gaussian longitudinal data. For continuous responses, we employ generalized linear models (GLM) for the marginals, while for binary or ordinal responses, we link factor copulas with an underlying normally distributed latent variable. This framework enables the incorporation of subject-specific covariates, essential components in longitudinal data analysis. We focus on univariate analysis of longitudinal responses of varying natures, assuming responses are missing completely at random. Model parameters are estimated using the IFM method (inference function of margins). Factor copula models can be juxtaposed with random effect models, such as generalized linear mixed models (GLMMs), as both involve latent variables at the subject level. Another noteworthy aspect of our study is the comparison of the performances of factor copula models with random effect models, despite their potentially divergent interpretations in different contexts.
\par The article is structured as follows: Section \ref{sec2} details factor copula constructions for (i) continuous responses using generalized linear models and (ii) discrete responses employing latent variables. Section \ref{sec3} elaborates on the estimation techniques and computational intricacies involved. Competing random effect models and commonly used model selection techniques are discussed in Section \ref{sec4}. In Section \ref{sec5}, diagnostics of the residuals of fitted factor copula models are presented. Extensive simulation studies to evaluate the finite sample performances of the proposed longitudinal models under parsimonious specifications of factor copulas are outlined in Section \ref{sec6}. Two motivating real-world datasets are described in Section \ref{sec7}, followed by the presentation of our analysis. Finally, Section \ref{sec8} concludes the article and outlines potential future extensions of the proposed models.
\section{The factor copula models for continuous and discrete responses}\label{sec2}
By the definition of longitudinal data, the repeated measurement tend to be serially correlated, hence for moderate to high-dimensional situations the use of factor copulas can be quite effective in terms of parametric inference. We assume that for the $i$-th subject the responses $\mathbf{Y}_i = (Y_{i1},\dots,Y_{in_i})^\intercal$ are conditionally independent given $p$ latent variables $V_{i1},\dots,V_{ip}$ which are independent and identically distributed (i.i.d) $U(0,1)$ (\citet{krupskii2013factor}). That is, a factor copula model is a conditional independence model which explains the dependence among the repeated measurements through some latent variables. The responses can be either continuous or discrete in nature. Here we describe the constructions of $1$-factor and $2$-factor copula models in the context of longitudinal data, but these can be extended up to $p (\geq 3)$ factors.
\par Consider the marginal density of a continuous response $Y_{ij}$ at the $j$-th measurement to be $f(y_{ij}|\theta_{ij})$, where $\theta_{ij}$ is a parameter which may depend on a vector of covariates $\mathbf{x}_{ij}$. For a $1$-factor copula model, let $V_{i1}$ be an $U(0,1)$ variable. Here the suffix `$i$' stands for a subject, and hence $V_{i1}$ and $V_{i^*1}$ are considered to be independent ($i \neq i^*$) in a longitudinal setting. The joint distribution of $Y_{ij}$ and $V_{i1}$ can be written in terms of a copula $C_{j,1}(.|\phi_{ij})$ as 
\begin{equation}\label{bicop1}
P(Y_{ij} \leq y_{ij},V_{i1} \leq v_{i1}) = C_{j,1}(F_{ij}(y_{ij}|\theta_{ij}),v_{i1}|\phi_{ij}),
\end{equation}
where $F_{ij}(.|\theta_{ij})$ is the cumulative distribution function of $Y_{ij}$ and $\phi_{ij}$ is the dependence parameter associated with the copula $C_{j,1}$. Therefore, we have the conditional distribution of $Y_{ij}$ given $V_{i1}$ as
\begin{equation}\label{bicop2}
F_{ij|1}(y_{ij}|v_{i1},\theta_{ij},\phi_{ij}) = \frac{\partial C_{j,1}(F_{ij}(y_{ij}|\theta_{ij}),v_{i1}|\phi_{ij})}{\partial v_{i1}} = C_{j|1}(F_{ij}(y_{ij}|\theta_{ij})|v_{i1},\phi_{ij}). 
\end{equation}
Then the joint distribution of $\mathbf{Y}_i$ is obtained as
\begin{equation}\label{c1ftdis1}
F_{n_i}(y_{i1},\dots,y_{in_i}|\mathbf{\theta}^*_i) = \int_0^1 \prod_{j=1}^{n_i} F_{ij|1}(y_{ij}|v_{i1},\theta_{ij},\phi_{ij}) dv_{i1} = \int_0^1 \prod_{j=1}^{n_i} C_{j|1}(F_{ij}(y_{ij}|\theta_{ij})|v_{i1},\phi_{ij}) dv_{i1}
\end{equation}
with the corresponding density function -
\begin{equation}\label{c1ftden1}
f_{n_i}(y_{i1},\dots,y_{in_i}|\mathbf{\theta}^*_i) = \int_0^1 \prod_{j=1}^{n_i} c_{j,1}(F_{ij}(y_{ij}|\theta_{ij}),v_{i1}|\phi_{ij}) dv_{i1} \prod_{j=1}^{n_i} f_{ij}(y_{ij}|\theta_{ij})
\end{equation}
where $\mathbf{\theta}^*_i = (\mathbf{\theta}_i,\mathbf{\phi}_i)^\intercal$ contains all the marginal and dependence parameters for the $i$-th subject and $c_{j,1}$ is the copula density function. Thus in a $1$-factor copula model $n_i$ bivariate copulas couple each observed variable to the first latent variable. For a $2$-factor copula model we consider $2$ latent variables $V_{i1}$ and $V_{i2}$, i.i.d $\sim U(0,1)$. The joint distribution of $Y_{ij}$ and $V_{i2}$ given $V_{i1}$, can be written in terms of a copula $C_{j,2}(.|\delta_{ij})$ as
\begin{equation}\label{bicop3}
P(Y_{ij} \leq y_{ij},V_{i2} \leq v_{i2}|V_{i1} = v_{i1}) = C_{j,2}(F_{ij|1}(y_{ij}|v_{i1},\theta_{ij},\phi_{ij}),v_{i2}|\delta_{ij}), 
\end{equation}
where $\delta_{ij}$ is the dependence parameter associated with the copula $C_{j,2}$. Similarly, the conditional distribution of $Y_{ij}$ given $V_{i1}$ and $V_{i2}$ is given as
\begin{align}\label{bicop4}
F_{ij|1,2}(y_{ij}|v_{i1},v_{i2},\theta_{ij},\phi_{ij},\delta_{ij}) & = \frac{\partial C_{j,2}(F_{ij|1}(y_{ij}|v_{i1},\theta_{ij},\phi_{ij}),v_{i2}|\delta_{ij})}{\partial v_{i2}} \nonumber \\ & = C_{j|1,2}(F_{ij|1}(y_{ij}|v_{i1},\theta_{ij},\phi_{ij})|v_{i2},\delta_{ij}).
\end{align}
Therefore, the joint distribution function of $\mathbf{Y}_i$ is obtained as
\begin{align}\label{c2ftdis1}
F_{ni}(y_{i1},\dots,y_{in_i}|\mathbf{\theta}^*_i) & = \int_0^1 \int_0^1 \prod_{j=1}^{n_i} F_{ij|1,2}(y_{ij}|v_{i1},v_{i2},\theta_{ij},\phi_{ij},\delta_{ij}) dv_{i1} dv_{i2} \nonumber \\ & = \int_{0}^1 \int_0^1 \prod_{j=1}^{n_i} C_{j|1,2}(F_{ij|1}(y_{ij}|v_{i1},\theta_{ij},\phi_{ij})|v_{i2},\delta_{ij}) dv_{i1} dv_{i2}
\end{align}
and the density function, $f_{ni}(y_{i1},\dots,y_{in_i}|\mathbf{\theta}^*_i) =$
\begin{equation}\label{c2ftden1}
\int_0^1 \int_0^1 \prod_{j=1}^{n_i} c_{j,2}(F_{ij|1}(y_{ij}|v_{i1},\theta_{ij},\phi_{ij}),v_{i2}|\delta_{ij}) c_{j,1}(F_{ij}(y_{ij}|\theta_{ij}),v_{i1}|\phi_{ij}) dv_{i1} dv_{i2} \prod_{j=1}^{n_i} f_{ij}(y_{ij}|\theta_{ij}).     
\end{equation}
In a $2$-factor copula model, there are another set of $n_i$ bivariate copulas that link each observed variable to the second latent variable conditioned on the first factor.
\par For discrete longitudinal responses (binary or ordinal), copulas are indirectly applied through some latent variables. Generally these variables are assumed to be normally distributed (\citet{agresti2010analysis}). Let $Y_{ij}$ represent a categorical response with $K$ possible ordered categories and let $Z_{ij}$ be a normally distributed latent variable underneath $Y_{ij}$. Let $\gamma(k), 1 < k < K - 1$, be ordered thresholds such that: $ -\infty = \gamma(0) < \gamma(1) < \cdots < \gamma(K -1) < \gamma(K) = \infty$. Then the discrete response have the stochastic representation as
\begin{equation}\label{disrep1}
Y_{ij} = k \;\; \text{if} \;\; \gamma(k - 1) \leq Z_{ij} < \gamma(k), \;\; k \in \{1,\dots,K\}. 
\end{equation}
The threshold parameters can be fixed or freely estimated based on the specification of the model. Similar to the continuous case the proxy variables $Z_{ij}$ may depend on a vector of covariates $\mathbf{x}_{ij}$. For $1$-factor copula model the conditional distribution of $Y_{ij}$ given $V_{i1}$ is therefore
\begin{align}\label{bicop5}
P(Y_{ij} = y_{ij}|V_{i1} = v_{i1}) & = P(Z_{ij} < \gamma(y_{ij})|V_{i1} = v_{i1}) - P(Z_{ij} < \gamma(y_{ij} - 1)|V_{i1} = v_{i1}) \nonumber \\ & = F_{ij|1}(\gamma(y_{ij})|v_{i1},\theta_{ij},\phi_{ij}) - F_{ij|1}(\gamma(y_{ij} - 1)|v_{i1},\theta_{ij},\phi_{ij}) \nonumber \\ & = C_{j|1}(F_{ij}(\gamma(y_{ij})|\theta_{ij})|v_{i1},\phi_{ij}) - C_{j|1}(F_{ij}(\gamma(y_{ij} - 1)|\theta_{ij})|v_{i1},\phi_{ij}).
\end{align}
Hence the joint distribution (pmf) of $\mathbf{Y}_i$ is obtained as
\begin{align}\label{c1ftdis2}
& f_{n_i}(y_{i1},\dots,y_{in_i}|\mathbf{\theta}^*_i) \nonumber \\ & = \int_0^1 \prod_{j=1}^{n_i} P(Y_{ij} = y_{ij}|V_{i1} = v_{i1}) dv_{i1} \nonumber \\ & = \int_0^1 \prod_{j=1}^{n_i} \Big[C_{j|1}(F_{ij}(\gamma(y_{ij})|\theta_{ij})|v_{i1},\phi_{ij}) - C_{j|1}(F_{ij}(\gamma(y_{ij} - 1)|\theta_{ij})|v_{i1},\phi_{ij})\Big] dv_{i1}.
\end{align}
For a $2$-factor copula model the conditional distribution of $Y_{ij}$ given $V_{i1}$ and $V_{i2}$ is given by -
\begin{align}\label{bicop6}
& P(Y_{ij} = y_{ij}|V_{i1} = v_{i1},V_{i2} = v_{i2}) \nonumber \\ & = P(Z_{ij} < \gamma(y_{ij})|V_{i1} = v_{i1},V_{i2} = v_{i2}) - P(Z_{ij} < \gamma(y_{ij} - 1)|V_{i1} = v_{i1},V_{i2} = v_{i2}) \nonumber \\ & = C_{j|1,2}(F_{ij|1}(\gamma(y_{ij})|v_{i1},\theta_{ij},\phi_{ij})|v_{i2},\delta_{ij}) - C_{j|1,2}(F_{ij|1}(\gamma(y_{ij} - 1)|v_{i1},\theta_{ij},\phi_{ij})|v_{i2},\delta_{ij}).
\end{align}
Hence the joint distribution (pmf) of $\mathbf{Y}_i$ is obtained as
\begin{multline}\label{c2ftdis2}
f_{n_i}(y_{i1},\dots,y_{in_i}|\mathbf{\theta}^*_i) = \int_0^1 \int_0^1 \prod_{j=1}^{n_i} P(Y_{ij} = y_{ij}|V_{i1} = v_{i1},V_{i2} = v_{i2}) dv_{i1} dv_{i2} \\ = \int_0^1 \int_0^1 \prod_{j=1}^{n_i} \Big[C_{j|1,2}(F_{ij|1}(\gamma(y_{ij})|v_{i1},\theta_{ij},\phi_{ij})|v_{i2},\delta_{ij}) - C_{j|1,2}(F_{ij|1}(\gamma(y_{ij} - 1)|v_{i1},\theta_{ij},\phi_{ij})|v_{i2},\delta_{ij})\Big] dv_{i1} dv_{i2}.
\end{multline}
Here we are making the so called simplifying assumption that the conditional copula for the univariate distributions $F_{ij|1}$ and $v_{i2}$ does not depend on $v_{i1}$. The choice of bivariate linking copulas in $1$-factor and $2$-factor copula models is completely arbitrary, but in this article we restrict to the elliptical copulas in the sense that the dependence parameter have direct interpretation same as their bivariate distributions. This is also motivated from the distributional properties of elliptical copulas. We consider the bivariate Gaussian copula as
\begin{equation}\label{gauss1}
C(u_1,u_2|\rho) = \Phi_2(\Phi^{-1}(u_1),\Phi^{-1}(u_2)|\rho), \quad -1 \leq \rho \leq 1,
\end{equation}
where $\Phi_2$ is the bivariate standard normal cdf with correlation parameter $\rho$, and $\Phi^{-1}$ is the quantile of the univariate standard normal. \citet{krupskii2013factor} and \citet{nikoloulopoulos2015factor} showed that if all bivariate copulas are normal with all normal marginals then factor copula models are multivariate normal and normal ogive models for the continuous and discrete case, respectively. We also consider the bivariate Student-$t$ copula as
\begin{equation}\label{st1}
C(u_1,u_2|\rho,\nu) = T_2(T^{-1}(u_1|\nu),T^{-1}(u_2|\nu)|\rho,\nu), \quad -1 \leq \rho \leq 1,
\end{equation}
where $T^{-1}$ is the quantile of the univariate Student-$t$ with $\nu$ degrees of freedom, and $T_2$ is the cdf of a bivariate Student-$t$ distribution with $\nu$ degrees of freedom and correlation parameter $\rho$. This additional degrees of freedom parameter $\nu$, accounts for possible tail dependence in the data. Small values of $\nu$, such as $2 \leq 5$, leads to a model with more probabilities in the joint upper and joint lower tails compared with the normal copula.
\par The final component of the specification of the factor copula based models is the choice of the marginals. Following a regression setting, we consider generalized linear models (GLM) for the marginals for the continuous responses (\citet{mcculloch2003generalized}). Let $\mathbf{x}_{ij}$ be a $p$ dimensional vector of covariates, $\mathbf{\beta}$ be a $p \times 1$ vector of regression coefficients and $g(.)$ be a model specific known link function then a generalized linear model can be expressed in the form:
\begin{equation}\label{glm1}
g(E(Y_{ij})) = \mathbf{x}_{ij}\mathbf{\beta}, \quad j = 1,\dots,n_i.
\end{equation}
For the discrete responses we model the latent variables based on covariate vector $\mathbf{x}_{ij}$ as
\begin{equation}\label{lat1}
Z_{ij} = \mathbf{x}_{ij}\mathbf{\beta} + \epsilon_{ij}, \quad j = 1,\dots,n_i,
\end{equation}
where the error term $\epsilon_{ij} \sim N(0,1)$. To ensure identifiaility of the ordinal model, we further fix the intercept parameter of $\mathbf{\beta}$ equal to zero. 
\section{Parameter estimation}\label{sec3}
The two most common frequentist technique of parameter estimation of copula based models is (i) maximum likelihood estimation (MLE), and (ii) inference function of margins (IFM) (\citet{joe1996estimation} and \citet{joe2005asymptotic}). The former is known to be the most efficient but can be computationally demanding for most the dependence models such as factor copula models. Here we implement the IFM method in which the model parameters are estimated in two separate stages. Note that ignorability property does applies for pseudo likelihood estimation (see, \citet{molenberghs2011pseudo}). First we estimate all marginal parameters assuming independence, by maximizing pseudo likelihood of the form: 
\begin{equation}\label{ifm1}
l(\mathbf{\theta}|\mathbf{y},\mathbf{x}) = \sum_{i=1}^m \sum_{j=1}^{n_i} \log f_{ij}(y_{ij}|\theta_{ij}),
\end{equation}
and then using the parameter estimates from (\ref{ifm1}), we compute the uniform samples: $u_{ij} = F_{ij}(y_{ij}|\hat{\theta}_{ij})$ for the continuous case, and $u_{ij} = F_{ij}(\gamma(y_{ij})|\hat{\theta}_{ij})$, $u_{ij}^- = F_{ij}(\gamma(y_{ij} - 1)|\hat{\theta}_{ij})$ for the discrete case, respectively where $i = 1,\dots,m, j = 1,\dots,n_i$. Thereafter we estimate the parameters of $1$-factor copula models for the continuous case by maximizing the pseudo likelihood:
\begin{equation}\label{ifm2}
l(\mathbf{\phi}|\mathbf{u}) = \sum_{i=1}^{m} \log \int_0^1 \prod_{j=1}^{n_i} c_{j,1}(u_{ij},v_{i1}|\phi_{ij}) dv_{i1},
\end{equation}
and for the discrete case by maximizing
\begin{equation}\label{ifm3}
l(\mathbf{\phi}|\mathbf{u}) = \sum_{i=1}^m \log \int_0^1 \prod_{j=1}^{n_i} \Big[ C_{j|1}(u_{ij}|v_{i1},\phi_{ij}) - C_{j|1}(u_{ij}^-|v_{i1},\phi_{ij}) \Big] dv_{i1}.
\end{equation}
Based on the parametric copula family we can estimate the conditional copulas using the corresponding h-functions. For the bivariate Gaussian copula this is
\begin{equation}\label{hgauss1}
C_{j|1}(u_1|u_2,\rho) = \Phi\left(\frac{\Phi^{-1}(u_1) - \rho \Phi^{-1}(u_2)}{\sqrt{1 - \rho^2}}\right),
\end{equation}
and for the bivariate Student-$t$ copula we compute
\begin{equation}\label{hst1}
C_{j|1}(u_1|u_2,\rho,\nu) = T\left(\frac{T^{-1}(u_1|\nu) - \rho T^{-1}(u_2|\nu)}{\sqrt{\frac{(\nu + T^{-1}(u_2|\nu))^2 (1 - \rho^2)}{\nu + 1}}} \Bigg|\nu + 1\right).
\end{equation}
Similarly in the $2$-factor copula models for the continuous case we estimate the dependence parameters by maximizing
\begin{equation}\label{ifm4}
l(\mathbf{\phi},\mathbf{\delta}|\mathbf{u}) = \sum_{i=1}^{m} \log \int_0^1 \int_0^1 \prod_{j=1}^{n_i}  c_{j,2}(C_{j|1}(u_{ij}|v_{i1},\phi_{ij}),v_{i2}|\delta_{ij}) c_{j,1}(u_{ij},v_{i1}|\phi_{ij}) dv_{i1} dv_{i2},
\end{equation}
and for the discrete case by maximizing
\begin{multline}\label{ifm5}
l(\mathbf{\phi},\mathbf{\delta}|\mathbf{u}) = \sum_{i=1}^m \log \int_0^1 \int_0^1 \prod_{j=1}^{n_i} \Big[ C_{j|1,2}(C_{j|1}(u_{ij}|v_{i1},\phi_{ij})|v_{i2},\delta_{ij}) - C_{j|1,2}(C_{j|1}(u_{ij}^-|v_{i1},\phi_{ij})|v_{i2},\delta_{ij}) \Big] dv_{i1} dv_{i2}.
\end{multline}
To carry out the numerical integrations in (\ref{ifm2}), (\ref{ifm3}), (\ref{ifm4}) and (\ref{ifm5}) we use Gauss-Hermite quadrature rule with $15$ quadrature points. The standard errors of the parameter estimates $\hat{\mathbf{\theta}}^* = (\hat{\mathbf{\theta}},\hat{\mathbf{\phi}})^\intercal$ of the $1$-factor and $2$-factor copula models can be numerically obtained from the estimated sandwich information matrix (Godambe information matrix) as
\begin{equation}\label{godambe1}
J(\hat{\mathbf{\theta}}^*) = D(\hat{\mathbf{\theta}}^*)^\intercal M(\hat{\mathbf{\theta}}^*)^{-1} D(\hat{\mathbf{\theta}}^*)
\end{equation}
where $D(\hat{\mathbf{\theta}}^*)$ is a block diagonal matrix and $M(\hat{\mathbf{\theta}}^*)$ is a symmetric positive definite matrix. More details regarding estimation of the matrix can be found in \citet{joe1996estimation} or \citet{joe2014dependence}. To estimate the parameters we use {\em optim} (\citet{cortez2014modern}) function, and to estimate the information matrix associated with the parameter estimates we use {\em numderiv} (\citet{gilbert2009package}) function in R.
\section{Comparison with random effect models}\label{sec4}
Random effects are generally used to introduce subject-specific effects to the linear predictor of the models as well as within-subject dependency or temporal dependence (see, \citet{shi2014longitudinal}). Factor copula models can be compared with the random effect models since both of these models involves unobservable latent variables. This means that models of this type are necessarily based on not fully verifiable assumption. For a detailed overview, we refer to the readers to see, \citet{molenberghs2012enriched} and \citet{verbeke2010arbitrariness}. In this article we also consider random effect models for accounting the temporal dependence as competitors by extending the marginal models introduced in Section \ref{sec2}. Referring to (\ref{glm1}), we consider generalized linear mixed models as
\begin{equation}\label{glmm1}
g(E(Y_{ij}|\mathbf{b}_i)) = \mathbf{x}_{ij}\mathbf{\beta} + \mathbf{d}_{ij}\mathbf{b}_i, \quad j = 1,\dots,n_i,
\end{equation}
where $\mathbf{d}_{ij}$ is the corresponding random effects design vector. Similarly following \citet{hedeker1994random}, we extend the latent variable models in (\ref{lat1}) by
\begin{equation}\label{latm1}
Z_{ij} = \mathbf{x}_{ij}\mathbf{\beta} + \mathbf{d}_{ij}\mathbf{b}_i + \epsilon_{ij}, \quad j = 1,\dots,n_i.
\end{equation}
Let $\mathbf{y}_i = (y_{i1},\dots,y_{in_i})^\intercal$ be the vector of observed responses for the $i$-th subject, ($i = 1,\dots,m$) then the parameters of the random effect models can be estimated from the marginal likelihood as
\begin{equation}\label{randest1}
l(\mathbf{\theta}|\mathbf{y},\mathbf{x}) = \sum_{i=1}^m \int \prod_{j=1}^{n_i} f(y_{ij}|\mathbf{b}_i) h(\mathbf{b}_i) d\mathbf{b}_i,
\end{equation}
where $h(.)$ is the distribution of the random effects which is assumed to be normal. For simplicity and applicability to the real world data sets considered in this article we restrict with random intercept models only. An important aspect of analysis of longitudinal data is model choice, or determining the number of components that is most appropriate for a specific data set. For this purpose we consider the two most widely used tool as AIC (Akaike information criterion) and BIC (Bayesian information criterion), which penalize large number of parameters. These are defined by
\begin{equation}\label{modsel1}
AIC = -2l(\hat{\mathbf{\theta}}^*) + 2\dim(\hat{\mathbf{\theta}}^*), \quad BIC = -2l(\hat{\mathbf{\theta}}^*) + \log(m)\dim(\hat{\mathbf{\theta}}^*)
\end{equation}
where $\hat{\mathbf{\theta}}^*$ is the maximum likelihood estimates of the model parameters and $m$ is the sample size. Each criteria described above has two terms; the first term for measuring the goodness-of-fit, and the second term for penalizing model complexity. However, when we estimate the parameters using two-stage IFM method, we still use them as some close approximations of the actual ones (see, \citet{ko2019copula} and \citet{killiches2018ad}). The smaller values of AIC and BIC indiacate a better fitting model. We use these criteria in our simulations and data analysis for comparing different models. But these selection criteria can not provide if the best fitting model is good enough for the data. For copula based models we can further validate the fitting using residual plots. 
\section{Residual analysis}\label{sec5}
Here we show how Rosenblatt's transformation (\citet{rosenblatt1952remarks}) is readily applicable for factor copula models. \citet{masarotto2012gaussian} or \citet{hofert2014graphical} previously used this method to validate copula assumptions for multivariate models. This method transforms dependent random variables into independent uniform random variables in the unit interval. Therefore, can be used to validate the distributional assumption of different multivariate models. Consistent with the setting considered in Section \ref{sec2} let
\begin{align}\label{rosen}
w_{i1} & = F(y_{i1}) \nonumber \\ & \dots \nonumber \\ w_{in_i} & = F(y_{in_i}|y_{i(n_i-1)},\dots,y_{i1},\hat{\mathbf{\theta}}^*),   
\end{align}
which are realizations of $n_i$ uncorrelated uniform variables if the model is correctly specified. For the continuous case under 1-factor copula model we have
\begin{align}\label{1fcop1}
F(y_{ij}|y_{i(j-1)},\dots,y_{i1},\hat{\mathbf{\theta}}^*) & = \int_0^1 F(y_{ij}|y_{i(j-1)},\dots,y_{i1},v_{i1},\hat{\mathbf{\theta}}^*)dv_{i1} \nonumber \\ & = \int_0^1 F(y_{ij}|v_{i1},\hat{\mathbf{\theta}}^*)dv_{i1} \nonumber \\ & = \int_0^1 C_{j|1}(\hat{u}_{ij}|v_{i1},\hat{\mathbf{\phi}}_{ij})dv_{i1},
\end{align}
where $\hat{u}_{ij} = F_{ij}(y_{ij}|\hat{\theta}_{ij})$. Similarly under 2-factor copula model we have
\begin{align}\label{2fcop1}
F(y_{ij}|y_{i(j-1)},\dots,y_{i1},\hat{\mathbf{\theta}}^*) & = \int_0^1 \int_0^1 F(y_{ij}|y_{i(j-1)},\dots,y_{i1},v_{i1},v_{i2},\hat{\mathbf{\theta}}^*)dv_{i1}dv_{i2} \nonumber \\ & = \int_0^1 \int_0^1 F(y_{ij}|,v_{i1},v_{i2},\hat{\mathbf{\theta}}^*)dv_{i1}dv_{i2} \\ & = \int_0^1 \int_0^1 C_{j|1,2}(C_{j|1}(\hat{u}_{ij}|v_{i1},\hat{\phi}_{ij})|v_{i2},\hat{\delta}_{ij})dv_{i1}dv_{i2}.
\end{align}
For the discrete case following \citet{zucchini2009hidden} or \citet{masarotto2012gaussian}, we compute the pseudo residuals by $w^*_{ij} = (w_{ij} + w^-_{ij})/2$ where $w^-_{ij} = F(y^-_{ij}|y_{i(j-1)},\dots,y_{i1},\hat{\mathbf{\theta}}^*)$ for $j = 1,\dots,n_i$. Therefore, quantiles of the residuals can be plotted against their expected values to graphically visualize the goodness-of-fit of the model.
\section{Simulation studies}\label{sec6}
To assess the performance of the estimation methods for our proposed models, we conduct simulation studies that emulate the characteristics of the datasets under consideration. We generate simulated datasets based on the proposed factor copula models and monitor the parametric inference process. We consider two distinct sample sizes, $m = \{200,500\}$, and set the maximum number of longitudinal responses to $d = 10$. To replicate the unbalanced nature of the data, for each unit, we generate $n_i$ from a binomial distribution with $d$ trials and success probability $p = 0.8$ (for $i = 1,\dots,m$), thereby pruning the dataset. Given that the true parameters are known, we initially sample $\mathbf{U}_i$ (in a vectorized notation) from $n_i$-variate ($n_i \leq d$) $1$-factor and $2$-factor copulas (for $i = 1,\dots,m$). Subsequently, we employ Probability Integral Transform (PIT) on $\mathbf{U}_i$ to generate per-unit responses $\mathbf{Y}_i$. Since the dimension for each response differs, we assume the same bivariate copulas for each factor with exchangeable parameters, resulting in a reduction in the number of estimable parameters from the model. We generate the continuous responses from the following model - 
\begin{equation}\label{contsim1}
g(E(Y_{ij})) = \beta_0 + x_{i1}\beta_1 + x_{i2}\beta_2 + t_{ij}\beta_3, \quad j = 1,\dots,n_i,
\end{equation}
where we consider two response distributions as Gamma ($\log$-link) and normal (identity-link), respectively. We assign same values of the regression coefficients for both of this marginals as, $\beta_0 = 1.0, \beta_1 = -0.5,\beta_2 = 0.2,\beta_3 = 0.2$ and set the dispersion parameters to $\kappa = 0.3$ (shape parameter of Gamma) and $\phi = 1.0$ (dispersion parameter of normal). Using the same dependence structure, we generate the binary responses from the following model -
\begin{align}\label{binsim1}
Y_{ij} & = \Big\{\begin{array}{cc} 0 & \text{if} \;\; Z_{ij} \leq 0 \\ 1 & \text{if} \;\; Z_{ij} > 0 \end{array}, \nonumber \\ Z_{ij} & = \beta_0 + x_{i1}\beta_1 + x_{i2}\beta_2 + t_{ij}\beta_3 + \epsilon_{ij}, \quad j = 1,\dots,n_i,
\end{align}
where $\epsilon_{ij} (i.i.d) \sim N(0,1)$ (independent and identically distributed). Here we assign the regression coefficients as, $\beta_0 = -0.5,\beta_1 = -0.5,\beta_2 = 0.2,\beta_3 = 0.2$. Finally we generate the ordinal responses from the following model -
\begin{align}\label{ordsim1}
Y_{ij} & = k \;\; \text{if} \;\; \gamma(k-1) \leq Z_{ij} <\gamma(k), \;\; k = 1,\dots,4, \nonumber \\ Z_{ij} & = x_{i1}\beta_1 + x_{i2}\beta_2 + t_{ij}\beta_3 + \epsilon_{ij}, \quad j = 1,\dots,n_i,
\end{align}
where $\epsilon_{ij} (i.i.d) \sim N(0,1)$. Here we assign the regression coefficients as, $\beta_1 = -0.5,\beta_2 = 0.2,\beta_3 = 0.2$ and the threshold parameters $\gamma_1 = -1.0,\gamma_2 = 1.0,\gamma_3 = 3.0$, respectively. For all the above considered models we generate the fixed covariates as $x_{i1} \sim Ber(p = 0.5)$, $X_{i2} \sim Unif(3,8)$ and the time points $t_{ij} = j$, for $j = 1,\dots,n_i, i = 1,\dots,m$, respectively.
For the $1$-factor copula models we set the correlation parameter $\rho_1 = 0.5$ and for the $2$-factor copula models we set the two correlation parameters $\{\rho_1,\rho_2\} = \{0.5,0.5\}$. We set the degrees of freedom parameter $\nu = 4.0$ when the bivariate copulas are assumed to be Student-$t$. For each model we simulate $N = 500$ Monte Carlo data sets, to monitor the performance under IFM estimation.
\begin{table}[]
    \centering
    \begin{small}
    \rotatebox{90}{
    \begin{minipage}{1.08\textwidth}
    \scalebox{1.0}{
    \tabcolsep = 0.18cm
    \begin{tabular}{|c|c c|c c c c c|c c c c c|}
    \hline
    \multicolumn{3}{|c|}{} & \multicolumn{5}{c|}{\textbf{m} = 200} & \multicolumn{5}{c|}{\textbf{m} = 500} \\
    \hline
    \textbf{Model} & \textbf{Parameters} & \textbf{True Value} & Mean & Bias & SD & SE & RMSE & Mean & Bias & SD & SE & RMSE \\
    \hline
    & $\beta_0$ & 1.0 & 0.9895 & -0.0105 & 0.0952 & 0.0985 & 0.0957 & 0.9875 & -0.0125 & 0.0648 & 0.0624 & 0.0686 \\
    Gamma & $\beta_1$ & -0.5 & -0.5000 & 0.0000 & 0.0480 & 0.0473 & 0.0480 & -0.4964 & 0.0036 & 0.0314 & 0.0300 & 0.0316 \\
    & $\beta_2$ & 0.2 & 0.2014 & 0.0014 & 0.0155 & 0.0164 & 0.0156 & 0.2031 & 0.0031 & 0.0105 & 0.0103 & 0.0110 \\
    & $\beta_3$ & 0.2 & 0.2006 & 0.0006 & 0.0053 & 0.0053 & 0.0053 & 0.2006 & 0.0006 & 0.0035 & 0.0034 & 0.0035 \\
    & $\kappa$ & 3.0 & 3.0160 & 0.0160 & 0.1228 & 0.1244 & 0.1238 & 3.0158 & 0.0158 & 0.0731 & 0.0784 & 0.0748 \\
    & $\rho_1$ & 0.5 & 0.4959 & -0.0042 & 0.0274 & 0.0225 & 0.0277 & 0.4959 & -0.0041 & 0.0169 & 0.0142 & 0.0174 \\
    \hline
    & $\beta_0$ & 1.0 & 1.0152 & 0.0152 & 0.1739 & 0.1730 & 0.1745 & 1.0034 & 0.0034 & 0.1130 & 0.1099 & 0.1131 \\
    Normal & $\beta_1$ & -0.5 & -0.5034 & -0.0034 & 0.0841 & 0.0830 & 0.0842 & -0.4995 & 0.0005 & 0.0528 & 0.0525 & 0.0528 \\
    & $\beta_2$ & 0.2 & 0.1978 & -0.0022 & 0.0292 & 0.0287 & 0.0293 & 0.1995 & -0.0005 & 0.0190 & 0.0182 & 0.0190 \\
    & $\beta_3$ & 0.2 & 0.1998 & -0.0002 & 0.0091 & 0.0091 & 0.0091 & 0.1999 & -0.0001 & 0.0056 & 0.0058 & 0.0056 \\
    & $\phi$ & 1.0 & 0.9969 & -0.0032 & 0.0209 & 0.0208 & 0.0211 & 0.9979 & -0.0021 & 0.0132 & 0.0132 & 0.0133 \\
    & $\rho_1$ & 0.5 & 0.4958 & -0.0042 & 0.0278 & 0.0226 & 0.0281 & 0.4974 & -0.0026 & 0.0169 & 0.0143 & 0.0171 \\
    \hline
    & $\beta_0$ & -0.5 & -0.5114 & -0.0114 & 0.2390 & 0.2318 & 0.2392 & -0.5008 & -0.0008 & 0.1420 & 0.1462 & 0.1420 \\
    Binary & $\beta_1$ & -0.5 & -0.4968 & 0.0032 & 0.1063 & 0.1119 & 0.1064 & -0.5065 & -0.0065 & 0.0766 & 0.0708 & 0.0768 \\
    & $\beta_2$ & 0.2 & 0.2025 & 0.0025 & 0.0406 & 0.0391 & 0.0406 & 0.2016 & 0.0016 & 0.0234 & 0.0246 & 0.0235 \\
    & $\beta_3$ & 0.2 & 0.2011 & 0.0011 & 0.0185 & 0.0185 & 0.0186 & 0.2003 & 0.0003 & 0.0123 & 0.0117 & 0.0123 \\
    & $\rho_1$ & 0.5 & 0.4913 & -0.0087 & 0.0536 & 0.0518 & 0.0543 & 0.4949 & -0.0051 & 0.0336 & 0.0326 & 0.0340 \\
    \hline
    & $\beta_1$ & -0.5 & -0.5022 & -0.0020 & 0.0913 & 0.0889 & 0.0914 & -0.4976 & 0.0024 & 0.0597 & 0.0563 & 0.0598 \\
    Ordinal & $\beta_2$ & 0.2 & 0.1990 & -0.0010 & 0.0312 & 0.0310 & 0.0313 & 0.1998 & -0.0002 & 0.0198 & 0.0196 & 0.0198 \\
    & $\beta_3$ & 0.2 & 0.2004 & 0.0004 & 0.0117 & 0.0120 & 0.0117 & 0.2002 & 0.0002 & 0.0073 & 0.0076 & 0.0074 \\
    & $\gamma_1$ & -1.0 & -1.0258 & -0.0258 & 0.2197 & 0.2089 & 0.2212 & -1.0014 & -0.0014 & 0.1300 & 0.1318 & 0.1300 \\
    & $\gamma_2$ & 1.0 & 0.9919 & -0.0081 & 0.1907 & 0.1970 & 0.1908 & 1.0027 & 0.0027 & 0.1162 & 0.1097 & 0.1163 \\
    & $\gamma_3$ & 3.0 & 2.9964 & -0.0036 & 0.2024 & 0.2124 & 0.2025 & 3.0067 & 0.0067 & 0.1228 & 0.1210 & 0.1230 \\
    & $\rho_1$ & 0.5 & 0.4935 & -0.0065 & 0.0337 & 0.0318 & 0.0343 & 0.4977 & -0.0023 & 0.0200 & 0.0198 & 0.0202 \\
    \hline
    \end{tabular}}
    \caption{Parameter estimation using IFM method for Gaussian $1$-factor copula model with continuous and discrete marginals for $N = 500$ simulated data sets.}
    \label{tab:sim1ftgauss1}
    \end{minipage}}
    \end{small}
\end{table}
\begin{table}[]
    \centering
    \begin{small}
    \rotatebox{90}{
    \begin{minipage}{1.08\textwidth}
    \scalebox{1.0}{
    \tabcolsep = 0.18cm
    \begin{tabular}{|c|c c|c c c c c|c c c c c|}
    \hline
    \multicolumn{3}{|c|}{} & \multicolumn{5}{c|}{\textbf{m} = 200} & \multicolumn{5}{c|}{\textbf{m} = 500} \\
    \hline
    \textbf{Model} & \textbf{Parameters} & \textbf{True Value} & Mean & Bias & SD & SE & RMSE & Mean & Bias & SD & SE & RMSE \\
    \hline
    & $\beta_0$ & 1.0 & 0.9971 & -0.0029 & 0.1246 & 0.1182 & 0.1246 & 0.9935 & -0.0065 & 0.0757 & 0.0752 & 0.0775 \\
    Gamma & $\beta_1$ & -0.5 & -0.5022 & -0.0022 & 0.0604 & 0.0572 & 0.0605 & -0.4989 & 0.0011 & 0.0359 & 0.0365 & 0.0359 \\
    & $\beta_2$ & 0.2 & 0.2000 & 0.0000 & 0.0208 & 0.0199 & 0.0208 & 0.2003 & 0.0003 & 0.0125 & 0.0126 & 0.0127 \\
    & $\beta_3$ & 0.2 & 0.2001 & 0.0001 & 0.0050 & 0.0049 & 0.0050 & 0.2004 & 0.0004 & 0.0031 & 0.0031 & 0.0031 \\
    & $\kappa$ & 3.0 & 3.0410 & 0.0410 & 0.1553 & 0.1576 & 0.1606 & 3.0178 & 0.0178 & 0.0987 & 0.1008 & 0.1003 \\
    & $\rho_1$ & 0.5 & 0.5192 & 0.0192 & 0.0630 & 0.0651 & 0.0659 & 0.5182 & 0.0182 & 0.0508 & 0.0495 & 0.0540 \\
    & $\rho_2$ & 0.5 & 0.4432 & -0.0568 & 0.1129 & 0.1091 & 0.1146 & 0.4611 & -0.0389 & 0.0822 & 0.0775 & 0.0842 \\
    \hline
    & $\beta_0$ & 1.0 & 1.0133 & 0.0133 & 0.2111 & 0.2078 & 0.2111 & 1.0050 & 0.0050 & 0.1333 & 0.1321 & 0.1334 \\
    Normal & $\beta_1$ & -0.5 & -0.5099 & -0.0099 & 0.1000 & 0.1010 & 0.1005 & -0.5008 & -0.0008 & 0.0644 & 0.0642 & 0.0644 \\
    & $\beta_2$ & 0.2 & 0.1980 & -0.0020 & 0.0353 & 0.0349 & 0.0354 & 0.1995 & -0.0005 & 0.0221 & 0.0222 & 0.0221 \\
    & $\beta_3$ & 0.2 & 0.2002 & 0.0002 & 0.0087 & 0.0085 & 0.0087 & 0.1997 & -0.0003 & 0.0052 & 0.0054 & 0.0052 \\
    & $\phi$ & 1.0 & 0.9948 & -0.0052 & 0.0280 & 0.0264 & 0.0284 & 0.9985 & -0.0015 & 0.0170 & 0.0171 & 0.0170 \\
    & $\rho_1$ & 0.5 & 0.5206 & 0.0206 & 0.0625 & 0.0575 & 0.0658 & 0.5164 & 0.0164 & 0.0555 & 0.0512 & 0.0604 \\
    & $\rho_2$ & 0.5 & 0.4472 & -0.0528 & 0.1031 & 0.1003 & 0.1051 & 0.4612 & -0.0388 & 0.0889 & 0.0803 & 0.0927 \\
    \hline
    & $\beta_0$ & -0.5 & -0.5038 & -0.0038 & 0.2757 & 0.2649 & 0.2758 & -0.4972 & 0.0028 & 0.1778 & 0.1699 & 0.1780 \\
    Binary & $\beta_1$ & -0.5 & -0.4971 & 0.0029 & 0.1341 & 0.1310 & 0.1342 & -0.5000 & 0.0000 & 0.0890 & 0.0833 & 0.0890 \\
    & $\beta_2$ & 0.2 & 0.2005 & 0.0005 & 0.0471 & 0.0455 & 0.0471 & 0.1996 & -0.0004 & 0.0291 & 0.0291 & 0.0291 \\
    & $\beta_3$ & 0.2 & 0.2012 & 0.0012 & 0.0186 & 0.0184 & 0.0187 & 0.2002 & 0.0002 & 0.0113 & 0.0118 & 0.0113 \\
    & $\rho_1$ & 0.5 & 0.5176 & 0.0176 & 0.0756 & 0.0698 & 0.0778 & 0.5070 & 0.0070 & 0.0472 & 0.0443 & 0.0477 \\
    & $\rho_2$ & 0.5 & 0.4325 & -0.0675 & 0.1510 & 0.1501 & 0.1520 & 0.4795 & -0.0206 & 0.0848 & 0.0821 & 0.0851 \\
    \hline
    & $\beta_1$ & -0.5 & -0.5023 & -0.0023 & 0.1034 & 0.1065 & 0.1035 & -0.5010 & -0.0010 & 0.0685 & 0.0675 & 0.0685 \\
    Ordinal & $\beta_2$ & 0.2 & 0.2009 & 0.0009 & 0.0384 & 0.0370 & 0.0384 & 0.2018 & 0.0018 & 0.0233 & 0.0234 & 0.0234 \\
    & $\beta_3$ & 0.2 & 0.2018 & 0.0018 & 0.0119 & 0.0120 & 0.0120 & 0.2012 & 0.0012 & 0.0075 & 0.0076 & 0.0076 \\
    & $\gamma_1$ & -1.0 & -1.0152 & -0.0152 & 0.2490 & 0.2397 & 0.2495 & -0.9993 & 0.0007 & 0.1547 & 0.1532 & 0.1547 \\
    & $\gamma_2$ & 1.0 & 1.0127 & 0.0127 & 0.2355 & 0.2208 & 0.2358 & 1.0114 & 0.0114 & 0.1384 & 0.1403 & 0.1388 \\
    & $\gamma_3$ & 3.0 & 3.0241 & 0.0241 & 0.2501 & 0.2376 & 0.2513 & 3.0167 & 0.0167 & 0.1498 & 0.1512 & 0.1507 \\
    & $\rho_1$ & 0.5 & 0.4975 & -0.0028 & 0.0310 & 0.0298 & 0.0310 & 0.5036 & 0.0036 & 0.0213 & 0.0210 & 0.0226 \\
    & $\rho_2$ & 0.5 & 0.4917 & -0.0083 & 0.0387 & 0.0308 & 0.0388 & 0.4977 & -0.0023 & 0.0282 & 0.0273 & 0.0283 \\ 
    \hline
    \end{tabular}}
    \caption{Parameter estimation using IFM method for Gaussian $2$-factor copula model with continuous and discrete marginals for $N = 500$ simulated data sets.}
    \label{tab:sim2ftgauss1}
    \end{minipage}}
    \end{small}
\end{table}
\begin{table}[]
    \centering
    \begin{small}
    \rotatebox{90}{
    \begin{minipage}{1.08\textwidth}
    \scalebox{1.0}{
    \tabcolsep = 0.18cm
    \begin{tabular}{|c|c c|c c c c c|c c c c c|}
    \hline
    \multicolumn{3}{|c|}{} & \multicolumn{5}{c|}{\textbf{m} = 200} & \multicolumn{5}{c|}{\textbf{m} = 500} \\
    \hline
    \textbf{Model} & \textbf{Parameters} & \textbf{True Value} & Mean & Bias & SD & SE & RMSE & Mean & Bias & SD & SE & RMSE \\
    \hline
    & $\beta_0$ & 1.0 & 0.9920 & -0.0080 & 0.1064 & 0.0986 & 0.1067 & 0.9941 & -0.0059 & 0.0691 & 0.0626 & 0.0695 \\
    Gamma & $\beta_1$ & -0.5 & -0.5012 & -0.0012 & 0.0451 & 0.0472 & 0.0452 & -0.4981 & 0.0019 & 0.0296 & 0.0300 & 0.0296 \\
    & $\beta_2$ & 0.2 & 0.2014 & 0.0014 & 0.0172 & 0.0163 & 0.0173 & 0.2021 & 0.0021 & 0.0113 & 0.0104 & 0.0115 \\
    & $\beta_3$ & 0.2 & 0.1990 & -0.0010 & 0.0055 & 0.0052 & 0.0055 & 0.2003 & 0.0003 & 0.0035 & 0.0034 & 0.0035 \\
    & $\kappa$ & 3.0 & 3.0293 & 0.0293 & 0.1486 & 0.1545 & 0.1514 & 3.0163 & 0.0163 & 0.0960 & 0.0985 & 0.0973 \\
    & $\rho_1$ & 0.5 & 0.4877 & -0.0123 & 0.0293 & 0.0233 & 0.0317 & 0.4905 & -0.0095 & 0.0185 & 0.0167 & 0.0208 \\
    \hline
    & $\beta_0$ & 1.0 & 1.0058 & 0.0058 & 0.1775 & 0.1713 & 0.1776 & 1.0045 & 0.0045 & 0.1066 & 0.1097 & 0.1067 \\
    Normal & $\beta_1$ & -0.5 & -0.5067 & -0.0067 & 0.0846 & 0.0820 & 0.0848 & -0.5020 & -0.0020 & 0.0531 & 0.0528 & 0.0531 \\
    & $\beta_2$ & 0.2 & 0.1986 & -0.0014 & 0.0303 & 0.0284 & 0.0303 & 0.1991 & -0.0009 & 0.0179 & 0.0181 & 0.0179 \\
    & $\beta_3$ & 0.2 & 0.1995 & -0.0005 & 0.0094 & 0.0092 & 0.0094 & 0.2002 & 0.0002 & 0.0056 & 0.0058 & 0.0056 \\
    & $\phi$ & 1.0 & 0.9964 & -0.0036 & 0.0268 & 0.0267 & 0.0270 & 0.9998 & -0.0002 & 0.0181 & 0.0171 & 0.0181 \\
    & $\rho_1$ & 0.5 & 0.4884 & -0.0116 & 0.0295 & 0.0233 & 0.0317 & 0.4926 & -0.0074 & 0.0183 & 0.0146 & 0.0197 \\
    \hline
    & $\beta_0$ & -0.5 & -0.5283 & -0.0283 & 0.2423 & 0.2387 & 0.2439 & -0.5035 & -0.0035 & 0.1452 & 0.1515 & 0.1453 \\
    Binary & $\beta_1$ & -0.5 & -0.5074 & -0.0074 & 0.1242 & 0.1161 & 0.1244 & -0.5074 & -0.0074 & 0.0721 & 0.0738 & 0.0725 \\
    & $\beta_2$ & 0.2 & 0.2060 & 0.0060 & 0.0406 & 0.0405 & 0.0407 & 0.2009 & 0.0009 & 0.0252 & 0.0256 & 0.0252 \\
    & $\beta_3$ & 0.2 & 0.2028 & 0.0028 & 0.0192 & 0.0191 & 0.0194 & 0.2014 & 0.0014 & 0.0125 & 0.0122 & 0.0125 \\
    & $\rho_1$ & 0.5 & 0.4846 & -0.0155 & 0.0638 & 0.0566 & 0.0657 & 0.4953 & -0.0047 & 0.0363 & 0.0351 & 0.0366 \\
    \hline
    & $\beta_1$ & -0.5 & -0.5036 & -0.0036 & 0.0899 & 0.0887 & 0.0900 & -0.4987 & 0.0013 & 0.0557 & 0.0562 & 0.0557 \\
    Ordinal & $\beta_2$ & 0.2 & 0.2025 & 0.0025 & 0.0318 & 0.0308 & 0.0319 & 0.2002 & 0.0002 & 0.0187 & 0.0196 & 0.0187 \\
    & $\beta_3$ & 0.2 & 0.2009 & 0.0009 & 0.0126 & 0.0125 & 0.0127 & 0.2005 & 0.0005 & 0.0077 & 0.0079 & 0.0077 \\
    & $\gamma_1$ & -1.0 & -1.0128 & -0.0128 & 0.2257 & 0.2142 & 0.2260 & -1.0035 & -0.0035 & 0.1303 & 0.1352 & 0.1304 \\
    & $\gamma_2$ & 1.0 & 1.0100 & 0.0100 & 0.1907 & 0.1866 & 0.1909 & 1.0019 & 0.0019 & 0.1146 & 0.1084 & 0.1146 \\
    & $\gamma_3$ & 3.0 & 3.0193 & 0.0193 & 0.2099 & 0.2058 & 0.2108 & 3.0043 & 0.0043 & 0.1251 & 0.1211 & 0.1252 \\
    & $\rho_1$ & 0.5 & 0.4943 & -0.0057 & 0.0339 & 0.0313 & 0.0344 & 0.4985 & -0.0015 & 0.0214 & 0.0196 & 0.0215 \\
    \hline
    \end{tabular}}
    \caption{Parameter estimation using IFM method for Student-$t$ ($\nu = 4$) $1$-factor copula model with continuous and discrete marginals for $N = 500$ simulated data sets.}
    \label{tab:sim1ftst1}
    \end{minipage}}
    \end{small}
\end{table}
\begin{table}[]
    \centering
    \begin{small}
    \rotatebox{90}{
    \begin{minipage}{1.08\textwidth}
    \scalebox{1.0}{
    \tabcolsep = 0.18cm
    \begin{tabular}{|c|c c|c c c c c|c c c c c|}
    \hline
    \multicolumn{3}{|c|}{} & \multicolumn{5}{c|}{\textbf{m} = 200} & \multicolumn{5}{c|}{\textbf{m} = 500} \\
    \hline
    \textbf{Model} & \textbf{Parameters} & \textbf{True Value} & Mean & Bias & SD & SE & RMSE & Mean & Bias & SD & SE & RMSE \\
    \hline
    & $\beta_0$ & 1.0 & 0.9828 & -0.0172 & 0.1227 & 0.1188 & 0.1239 & 0.9852 & -0.0148 & 0.0761 & 0.0753 & 0.0776 \\
    Gamma & $\beta_1$ & -0.5 & -0.4988 & 0.0012 & 0.0609 & 0.0573 & 0.0609 & -0.4999 & 0.0001 & 0.0361 & 0.0364 & 0.0361 \\
    & $\beta_2$ & 0.2 & 0.2024 & 0.0024 & 0.0202 & 0.0199 & 0.0203 & 0.2021 & 0.0021 & 0.0127 & 0.0126 & 0.0129 \\
    & $\beta_3$ & 0.2 & 0.2003 & 0.0003 & 0.0050 & 0.0049 & 0.0050 & 0.2006 & 0.0006 & 0.0032 & 0.0031 & 0.0032 \\
    & $\kappa$ & 3.0 & 3.0418 & 0.0418 & 0.1835 & 0.1894 & 0.1882 & 3.0175 & 0.0175 & 0.1179 & 0.1211 & 0.1192 \\
    & $\rho_1$ & 0.5 & 0.4964 & -0.0036 & 0.0887 & 0.0737 & 0.0887 & 0.5033 & 0.0033 & 0.0651 & 0.0668 & 0.0652 \\
    & $\rho_2$ & 0.5 & 0.4400 & -0.0600 & 0.0897 & 0.0778 & 0.0898 & 0.4680 & -0.0320 & 0.0626 & 0.0660 & 0.0630 \\
    \hline
    & $\beta_0$ & 1.0 & 1.0226 & 0.0226 & 0.1970 & 0.2072 & 0.1983 & 0.9981 & -0.0019 & 0.1368 & 0.1316 & 0.1368 \\
    Normal & $\beta_1$ & -0.5 & -0.5057 & -0.0057 & 0.1046 & 0.1000 & 0.1048 & -0.5015 & -0.0015 & 0.0642 & 0.0636 & 0.0643 \\
    & $\beta_2$ & 0.2 & 0.1970 & -0.0030 & 0.0332 & 0.0347 & 0.0333 & 0.2006 & 0.0006 & 0.0235 & 0.0220 & 0.0235 \\
    & $\beta_3$ & 0.2 & 0.1997 & 0.0003 & 0.0085 & 0.0085 & 0.0085 & 0.1997 & -0.0003 & 0.0054 & 0.0054 & 0.0054 \\
    & $\phi$ & 1.0 & 0.9947 & -0.0053 & 0.0336 & 0.0324 & 0.0340 & 0.9991 & -0.0009 & 0.0209 & 0.0208 & 0.0209 \\
    & $\rho_1$ & 0.5 & 0.4973 & -0.0027 & 0.0854 & 0.0733 & 0.0854 & 0.5108 & 0.0108 & 0.0689 & 0.0521 & 0.0698 \\
    & $\rho_2$ & 0.5 & 0.4502 & -0.0498 & 0.0897 & 0.0763 & 0.0898 & 0.4714 & -0.0286 & 0.0620 & 0.0544 & 0.0629 \\
    \hline
    & $\beta_0$ & -0.5 & -0.5050 & -0.0050 & 0.2788 & 0.2756 & 0.2789 & -0.4989 & 0.0011 & 0.1709 & 0.1729 & 0.1711 \\
    Binary & $\beta_1$ & -0.5 & -0.5121 & -0.0121 & 0.1500 & 0.1367 & 0.1504 & -0.5094 & -0.0094 & 0.0867 & 0.0864 & 0.0872 \\
    & $\beta_2$ & 0.2 & 0.2027 & 0.0027 & 0.0491 & 0.0477 & 0.0491 & 0.1990 & 0.0010 & 0.0296 & 0.0302 & 0.0297 \\
    & $\beta_3$ & 0.2 & 0.2019 & 0.0019 & 0.0204 & 0.0193 & 0.0205 & 0.2005 & 0.0005 & 0.0127 & 0.0125 & 0.0127 \\
    & $\rho_1$ & 0.5 & 0.4963 & -0.0037 & 0.0807 & 0.0791 & 0.0808 & 0.4983 & -0.0012 & 0.0603 & 0.0601 & 0.0604 \\
    & $\rho_2$ & 0.5 & 0.4230 & -0.0770 & 0.1599 & 0.1529 & 0.1617 & 0.4347 & -0.0653 & 0.1245 & 0.1311 & 0.1310 \\
    \hline
    & $\beta_1$ & -0.5 & -0.5074 & -0.0074 & 0.1078 & 0.1057 & 0.1080 & -0.5044 & -0.0044 & 0.0691 & 0.0671 & 0.0693 \\
    Ordinal & $\beta_2$ & 0.2 & 0.2008 & 0.0008 & 0.0355 & 0.0367 & 0.0355 & 0.2010 & 0.0010 & 0.0227 & 0.0234 & 0.0227 \\
    & $\beta_3$ & 0.2 & 0.2009 & 0.0009 & 0.0122 & 0.0126 & 0.0122 & 0.2004 & 0.0004 & 0.0082 & 0.0080 & 0.0082 \\
    & $\gamma_1$ & -1.0 & -1.0433 & -0.0433 & 0.2554 & 0.2490 & 0.2590 & -1.0097 & -0.0097 & 0.1574 & 0.1577 & 0.1577 \\
    & $\gamma_2$ & 1.0 & 0.9943 & -0.0057 & 0.2163 & 0.2205 & 0.2164 & 1.0020 & 0.0020 & 0.1418 & 0.1398 & 0.1418 \\
    & $\gamma_3$ & 3.0 & 3.0063 & 0.0063 & 0.2385 & 0.2427 & 0.2356 & 3.0045 & 0.0045 & 0.1559 & 0.1543 & 0.1560 \\
    & $\rho_1$ & 0.5 & 0.4912 & -0.0088 & 0.0463 & 0.0419 & 0.0463 & 0.4925 & -0.0075 & 0.0216 & 0.0199 & 0.0219 \\
    & $\rho_2$ & 0.5 & 0.4831 & -0.0169 & 0.0375 & 0.0298 & 0.0377 & 0.4955 & -0.0045 & 0.0265 & 0.0249 & 0.0267 \\ 
    \hline
    \end{tabular}}
    \caption{Parameter estimation using IFM method for Student-$t$ ($\nu = 4$) $2$-factor copula model with continuous and discrete marginals for $N = 500$ simulated data sets.}
    \label{tab:sim2ftst1}
    \end{minipage}}
    \end{small}
\end{table}
\par In Table \ref{tab:sim1ftgauss1}, \ref{tab:sim2ftgauss1}, \ref{tab:sim1ftst1} and \ref{tab:sim2ftst1}, we present the averages of the parameter estimates (denoted as Mean), the biases $[\frac{1}{N}\sum_{i=1}^N (\hat{\mathbf{\theta}}^*_j - \mathbf{\theta}^*)]$, empirical standard deviations (denoted as SD), average standard errors obtained from the asymptotic covariance matrices (denoted as SE) and roots of mean square errors $[\sqrt{\frac{1}{N}\sum_{i=1}^N (\hat{\mathbf{\theta}}^*_j - \mathbf{\theta}^*)^2}]$, where $\hat{\mathbf{\theta}}^*_j$ is the parameter estimates for the $j$-th sample. The results show consistent performance of the proposed models with IFM estimation as the biases and roots of the mean square errors decreases with increasing sample size. The average standard errors are pretty close to the empirical standard deviations for almost all the parameters in every models, which ensures the reliability of inference. The standard errors of the parameter estimates for the $2$-factor copula based models are slightly larger than of the $1$-factor copula based models. Binary responses provide least amount of information for parameter estimation, as we see the standard errors of the regression coefficients for the binary responses are largest among other response models. Overall our simulations successfully demonstrated the capabilities of factor copula models in moderate to high dimensions under unbalanced data structure. For specific applications, bivariate linking copulas other than the elliptical ones can be considered to construct the joint probability distribution.
\section{Applications}\label{sec7}
In this section, we introduce two datasets derived from real-world longitudinal studies featuring mixed-type responses. These datasets are readily accessible in the R packages {\em mixAK} and {\em lcmm}, respectively. However, our article concentrates on evaluating the temporal dependency of each longitudinal outcome (whether continuous or discrete) individually within a univariate framework.
\subsection{The PBC 910 data}
\begin{figure}
    \centering
    \includegraphics[width=15cm]{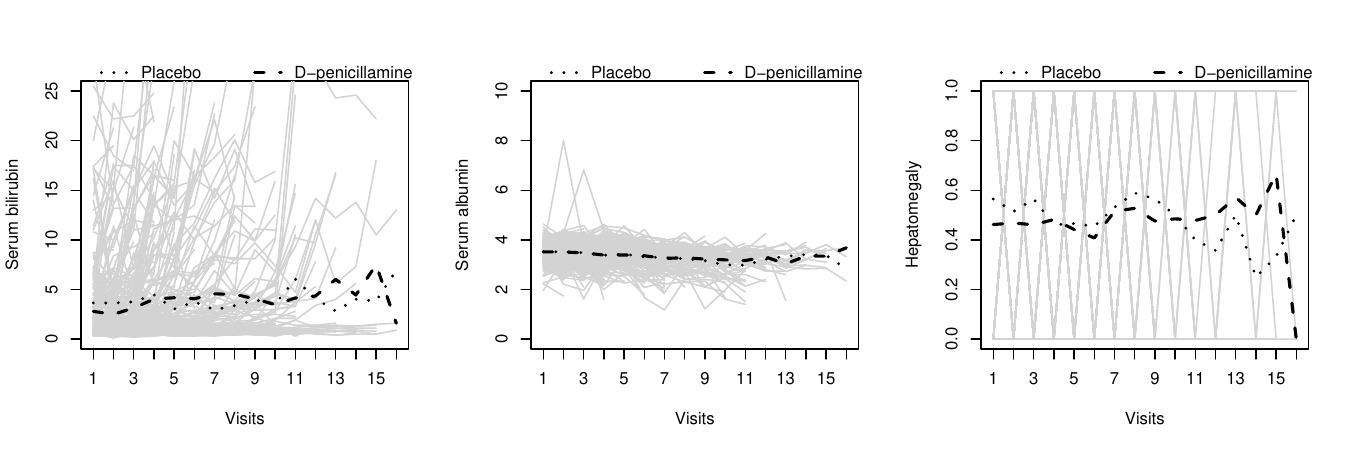}
    \caption{Subject-specific profiles over time for (i) Serum bilirubin, (ii) Serum albumin and (iii) Hepatomegaly for PBC 910 data set. The dotted lines show average profiles under placebo and D-penicillamine.}
    \label{fig:profile.pbc1}
\end{figure}

In clinical practices multiple markers of dicease progression are routinely gathered during the follow-up time to decide on future treatment actions. Motivated from the work of \citet{dickson1989prognosis}, we consider the laboratory data on patients with primary biliary cirrhosis (PBC), from a Mayo Clinic trial conducted in 1974-1984. PBC is a long-term liver disease in which the bile ducts in the liver become damaged. Progressively, this leads to a build-up of bile in the liver, which can damage it and eventually lead to cirrhosis. If PBC is not treated or reaches an advanced stage, it can lead to several major complications, including mortality. This longitudinal study had a median follow-up time of $6.3$ years, where $312$ patients were randomly assigned to receive D-penicillamine ($m = 158$) or placebo ($m = 154$). Several longitudinal biomarkers associated with liver function were serially recorded for these patients along with the baseline covariates such as gender and age. In this data set the number and timing of the measurements are quite different within and across subjects, resulting in a highly unbalanced longitudinal data. In the joint modeling literature, this data set has been analyzed by several authors (e.g. \citet{komarek2013clustering}, \citet{andrinopoulou2016bayesian} or \citet{hickey2018joinerml}). In our application we consider $3$ biomarkers: serum albumin (mg/dl), serum bilirubin (mg/dl) and hepatom (presence of hepatomegaly or enlarged liver). That is two continuous and one binary outcomes. Profile plots for each biomarker are shown in Figure \ref{fig:profile.pbc1}. It is seen that average profiles for both the control and D-penicillamine-treated patients are close to each other for each of the biomarkers, hence there are no theraputic differences between the two. Additionally, it is seen that, subjects have different number of visits reflecting the unbalancedness of the data. The biomarker Serum bilirubin is seen to be rightly skewed, hence it is not reasonable to assume the joint distribution over time as multivariate Gaussian. The maximum number of visit of the subjects in this data set is $16$, hence under the copula modeling framework it is difficult to address the temporal dependency with standard multivariate copulas.
\par For this data set we consider same set of covariates for each longitudinal responses. The fixed covariates are considered as sex ($0 =$ male, $1 =$ female), drug ($0 =$ placebo, $1 =$ D-penicillamine) and age along with the time of measurements (rescaled into years). For the two continuous responses we consider GLM for the marginals of the form -
\begin{equation}\label{creal1}
g(E(Y_{ij})) = \beta_0 + \text{sex}_i\beta_1 + \text{drug}_i\beta_2 + \text{age}_i\beta_3 + t_{ij}\beta_4, \quad j = 1,\dots,n_i, 
\end{equation}
where observed $y_{ij}$ is the continuous response at the $j$-th time for subject $i$. For the serum bilirubin marker, we consider Gamma marginals with $\log$-link and for the serum albumin marker we consider normal marginals with identity link, based on the graphical diagnostics. For the hepatom marker, we consider the latent variable model as
\begin{align}\label{breal1}
Y_{ij} & = \Big\{\begin{array}{cc} 0 & \text{if} \;\; Z_{ij} \leq 0 \\ 1 & \text{if} \;\; Z_{ij} > 0 \end{array}, \nonumber \\ Z_{ij} & = \beta_0 + \text{sex}_i\beta_1 + \text{drug}_i\beta_2 + \text{age}_i\beta_3 + t_{ij}\beta_4 + \epsilon_{ij}, \quad j = 1,\dots,n_i,
\end{align}
where $\epsilon_{ij} \sim N(0,1)$. Note that, unlike the D-vine copula approach in \citet{killiches2018ad}, the subjects with recorded measurements only once, still contributes to the density of the factor copula models. We consider the same parsimonious parameterization for the $1$-factor and $2$-factor copula models described in Section \ref{sec6}. We fit our models to each longitudinal responses assuming homogeneous dependence structure for all individuals, $i = 1,\dots,312$, over time. We also compare the fittings with respect to the corresponding random effect models as well.
\begin{table}[]
    \centering
    \begin{small}
    \scalebox{1.0}{
    \tabcolsep = 0.18cm
    \begin{tabular}{|c|c c|c|c c|c|c c|}
    \hline
    \multicolumn{3}{|c}{Serum bilirubin (gamma)} & \multicolumn{3}{|c}{Serum albumin (normal)} & \multicolumn{3}{|c|}{Hepatom (binary)} \\
    \hline
    \textbf{Parameters} & Est. & SE & \textbf{Parameters} & Est. & SE & \textbf{Parameters} & Est. & SE \\
    \hline
    $\beta_0$ & 1.8223 & 0.3799 & $\beta_0$ & 3.8159 & 0.1349 & $\beta_0$ & 0.1274 & 0.3174 \\
    $\beta_1$ & -0.2277 & 0.1745 & $\beta_1$ & -0.0194 & 0.0757 & $\beta_1$ & -0.3456 & 0.1495 \\
    $\beta_2$ & -0.0427 & 0.1372 & $\beta_2$ & 0.0369 & 0.0444 & $\beta_2$ & -0.1194 & 0.1066 \\
    $\beta_3$ & -0.0083 & 0.0062 & $\beta_3$ & -0.0061 & 0.0019 & $\beta_3$ & 0.0050 & 0.0054 \\
    $\beta_4$ & -0.0266 & 0.0198 & $\beta_4$ & -0.0401 & 0.0055 & $\beta_4$ & 0.0022 & 0.0158 \\
    $\kappa$ & 0.8596 & 0.0348 & - & - & - & - & - & - \\
    - & - & - & $\phi$ & 0.4845 & 0.0166 & - & - & - \\
    \hline
    \end{tabular}}
    \caption{Estimated marginal parameters and their standard errors of $3$ considered markers of the PBC$910$ data using the regression models in (\ref{creal1}) and (\ref{breal1}) respectively.}
    \label{tab:pbc910fit1}
    \end{small}
\end{table}
\begin{table}[]
    \centering
    \begin{small}
    \scalebox{1.0}{
    \tabcolsep = 0.18cm
    \begin{tabular}{|c|c|c|c|c|c|c|c|}
    \hline
    & \textbf{Copula} & \textbf{Parameters} & Est. & SE & Log-likelihood & AIC & BIC \\
    \hline
    & Gaussian & $\rho_1$ & 0.8234 & 0.0152 & -3703.78 & 7421.56 & 7447.76 \\
    Serum & $1$-factor & & & & & & \\
    bilirubin & Student-$t$ & $\rho_1$ & 0.8429 & 0.0125 & -3657.41 & 7330.82 & 7360.77 \\
    & $1$-factor ($\nu = 6$) & & & & & & \\
    & Gaussian & $\rho_1$ & 0.7234 & 0.0205 & -3606.78 & 7229.57 & 7259.52 \\
    & $2$-factor & $\rho_2$ & 0.6789 & 0.0349 & & & \\
    & Student-$t$ & $\rho_1$ & 0.8545 & 0.0544 & \textbf{-3353.31} & \textbf{6724.62} & \textbf{6758.31} \\ 
    & $2$-factor ($\nu = 3$) & $\rho_2$ & 0.7225 & 0.0619 & & & \\
    \hline
    & Gaussian & $\rho_1$ & 0.6663 & 0.0205 & -1052.75 & 2119.51 & 2145.71 \\
    Serum & $1$-factor & & & & & & \\
    albumin & Student-$t$ & $\rho_1$ & 0.6810 & 0.0187 & -1047.74 & 2111.48 & 2141.42 \\
    & $1$-factor ($\nu = 14$) & & & & & & \\
    & Gaussian & $\rho_1$ & 0.4831 & 0.0151 & -1043.97 & 2103.93 & 2133.88 \\
    & $2$-factor & $\rho_2$ & 0.5514 & 0.0226 & & & \\
    & Student-$t$ & $\rho_1$ & 0.4870 & 0.0183 & \textbf{-1032.35} & \textbf{2082.69} & \textbf{2116.38} \\
    & $2$-factor ($\nu = 9$) & $\rho_2$ & 0.5944 & 0.0245 & & & \\
    \hline
    & Gaussian & $\rho_1$ & 0.7283 & 0.0224 & -1140.27 & 2292.53 & 2314.99 \\
    Hepatom & $1$-factor & & & & & & \\
    & Student-$t$ & $\rho_1$ & 0.7296 & 0.0223 & -1140.69 & 2295.37 & 2321.58 \\
    & $1$-factor ($\nu = 30$) & & & & & & \\
    & Gaussian & $\rho_1$ & 0.5821 & 0.0340 & \textbf{-1137.54} & \textbf{2289.08} & \textbf{2315.28} \\
    & $2$-factor & $\rho_2$ & 0.5876 & 0.0368 & & & \\
    & Student-$t$ & $\rho_1$ & 0.5384 & 0.0212 & -1137.68 & 2291.37 & 2321.31 \\
    & $2$-factor ($\nu = 30$) & $\rho_2$ & 0.6282 & 0.0313 & & & \\
    \hline
    \end{tabular}}
    \caption{Estimated dependence parameters and their standard errors of $3$ considered markers of the PBC$910$ data with $1$-factor and $2$-factor copula models. Maximum log-likelihood value, AIC and BIC for each model are reported.}
    \label{tab:pbc910fit2}
    \end{small}
\end{table}
\begin{table}[]
    \centering
    \begin{small}
    \scalebox{1.0}{
    \tabcolsep = 0.18cm
    \begin{tabular}{|c|c c|c|c c|c|c c|}
    \hline
    \multicolumn{3}{|c}{Serum bilirubin (gamma)} & \multicolumn{3}{|c}{Serum albumin (normal)} & \multicolumn{3}{|c|}{Hepatom (binary)} \\
    \hline
    \textbf{Parameters} & Est. & SE & \textbf{Parameters} & Est. & SE & \textbf{Parameters} & Est. & SE \\
    \hline
    $\beta_0$ & 1.2441 & 0.5374 & $\beta_0$ & 3.9450 & 0.1373 & $\beta_0$ & 0.2093 & 0.4915 \\
    $\beta_1$ & -0.3437 & 0.2087 & $\beta_1$ & -0.0375 & 0.0707 & $\beta_1$ & -0.4465 & 0.2493 \\
    $\beta_2$ & 0.0056 & 0.1776 & $\beta_2$ & 0.0289 & 0.0448 & $\beta_2$ & -0.2412 & 0.1716 \\
    $\beta_3$ & -0.0039 & 0.0095 & $\beta_3$ & -0.0081 & 0.0022 & $\beta_3$ & 0.0060 & 0.0077 \\
    $\beta_4$ & 0.0900 & 0.0082 & $\beta_4$ & -0.0741 & 0.0030 & $\beta_4$ & 0.0505 & 0.0124 \\
    $V[b]$ & 1.0090 & 0.0987 & $V[b]$ & 0.1240 & 0.0125 & $V[b]$ & 1.4673 & 0.2155 \\
    $\kappa$ & 1.1926 & 0.0367 & - & - & - & - & - & - \\
    - & - & - & $\phi$ & 0.3510 & 0.0062 & - & - & - \\
    \hline
    \multicolumn{3}{|c}{Log-likelihood \quad \textbf{-3629.24}} & \multicolumn{3}{|c}{Log-likelihood \quad \textbf{-1033.84}} & \multicolumn{3}{|c|}{Log-likelihood \quad \textbf{-1125.45}} \\
    \multicolumn{3}{|c}{AIC \quad \textbf{7272.47}} & \multicolumn{3}{|c}{AIC \quad \textbf{2084.69}} & \multicolumn{3}{|c|}{AIC \quad \textbf{2262.90}} \\
    \multicolumn{3}{|c}{BIC \quad \textbf{7298.67}} & \multicolumn{3}{|c}{BIC \quad \textbf{2125.70}} & \multicolumn{3}{|c|}{BIC \quad \textbf{2285.36}} \\
    \hline
    \end{tabular}}
    \caption{Estimated parameters and their standard errors of $3$ considered markers of the PBC$910$ data by adding random intercepts to the regression models in (\ref{creal1}) and (\ref{breal1}) respectively. Maximum log-likelihood value, AIC and BIC for each model are reported.}
    \label{tab:pbc910fit3}
    \end{small}
\end{table}
\par
\begin{figure}
    \centering
    \includegraphics[width=15cm]{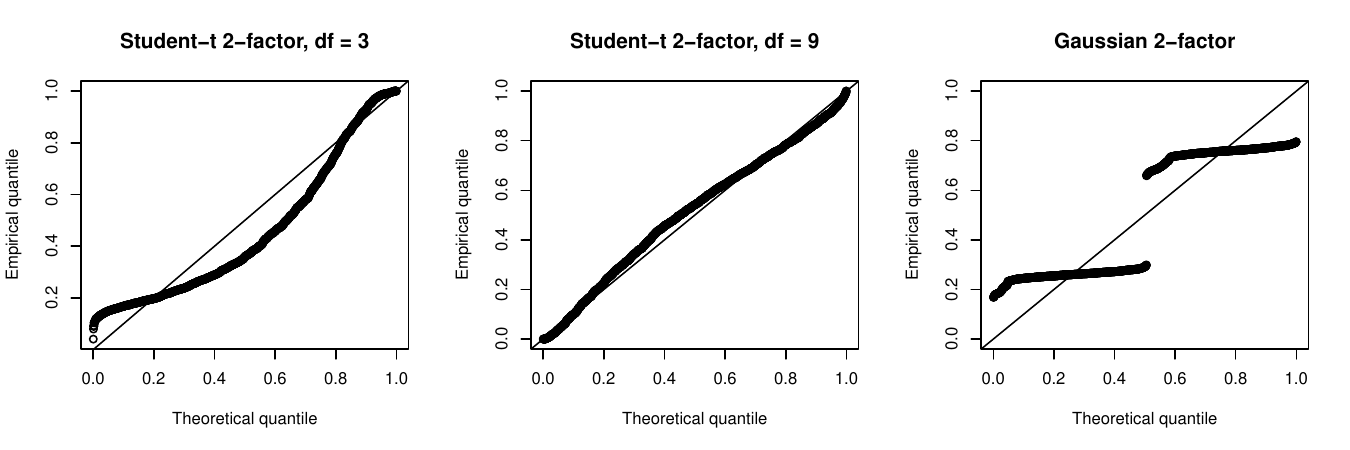}
    \caption{Uniform probability plots of the residuals of the best fitting copula models for (i) Serum bilirubin, (ii) Serum albumin and (iii) Hepatomegaly for PBC 910 data set.}
    \label{fig:resid.pbc1}
\end{figure}

 Table \ref{tab:pbc910fit1} and \ref{tab:pbc910fit2} presents the marginal parameter estimates and the dependence parameter estimates of the PBC$910$ data under different factor copula models, respectively. The estimates of $\beta_1$ for all the models are away from zero, which implies male subjects had higher progress in that liver disease than the female subjects. Both $1$-factor and $2$-factor copula models with Student-$t$ bivariate linking copulas outperformed the models with Gaussian bivariate linking copulas. Moreover, $2$-factor copula models provide with better fits than $1$-factor copula models with same bivariate linking copulas, which suggests one latent variable is not sufficient to describe the underlying dependence structure. The integer valued degrees of freedom parameter is obtained by the maximum value of the log-likelihood over the range of $\{3,\dots,30\}$. This indicates there are more probabilities in joint lower (or upper) tails of the bivariate pairs of each outcomes with the corresponding latent variables. The estimates of the copula parameters shows for each markers, there are strong correlations between the observed responses and the latent variables. Therefore, sticking with the elliptical copulas we can compare the dependence between the repeated measurements from the normal factor (\citet{krupskii2013factor}) and normal ogive (\citet{nikoloulopoulos2015factor}) models, which implies positive association between the repeated measurements of each markers. In Table \ref{tab:pbc910fit3} we provide the parameter estimates with random intercept models. The estimates of the regression coefficients are comparable except the intercept terms. For the serum albumin marker factor copula models outperforms the random effect models, but for the hepatom marker random effect models outperforms the factor copula models in terms of the selection criteria. For the serum albumin marker in terms of BIC, random effect models indicate better fits, but not in terms of AIC. We tried to fit these markers with random intercept and slope models as well but the likelihood did not converge for all the cases. In Figure \ref{fig:resid.pbc1} we provide the residual plots on uniform scale for each markers corresponding to their best fitting model. For the binary marker, since it is least informative no conclusion can be drawn from the residuals.
\subsection{The PAQUID data}
\begin{figure}
    \centering
    \includegraphics[width=15cm]{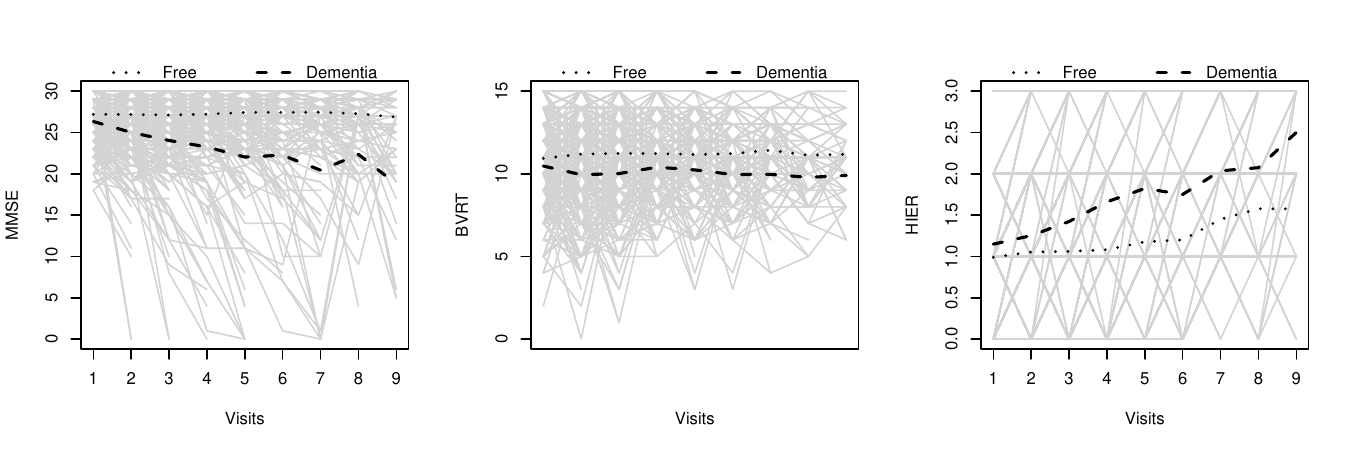}
    \caption{Subject-specific profiles over time for (i) MMSE, (ii) BVRT psychometric tests and (iii) HIER for PAQUID data set. The dotted lines show average profiles with free and positive diagnosis of dementia.}
    \label{fig:profile.paq1}
\end{figure}

We also consider the data set from French prospective cohort study PAQUID, initiated in $1988$ to study normal and pathological ageing (\citet{letenneur1994incidence}). This cohort included $3777$ individuals aged $65$ years and older at the initial visit and were followed six times with intervals of $2$ or $3$ years with repeated neuropsychological evaluations and clinical diagnoses of dementia (impaired ability to remember, think, or make decisions that interferes with doing everyday activities). At each visit, a battery of psychometric tests was completed and an evaluation of whether the person satisfied the criteria for a diagnosis of dementia was carried out. In our analysis, we consider $2$ psychometric tests as (i) the Mini-Mental State Examination (MMSE), which provides an index of global cognitive performance, (ii) the Benton Visual Retention Test (BVRT), which assesses visual memory and (iii) the score of physical dependency (HIER). Low values of the psychometric tests indicate a more severe impairment. This data set has also been analyzed under joint modeling setup in the literature (e.g. \citet{proust2006nonlinear}, \citet{proust2007nonlinear} or \citet{proust2019joint}]. Here we have two continuous and one ordinal outcomes. From the profile plots in Figure \ref{fig:profile.paq1}, we see that there are distinct differences in trajectories between those who diagnosed with dementia and those who did not. We consider a sub-sample of size $500$ in our analysis, with the aim of understanding evolution of each longitudinal responses. Out of those, $128$ had a positive diagnosis of dementia. These subjects had between $1$ to $9$ measurements per test with an average of $4.5$.
\par For this data set we consider the fixed covariates as sex ($1 =$ male, $0 =$ female), dem ($1 =$ diagnosed positive of dementia), cep (educational level, $1 =$ graduated from primary school, $0 =$ otherwise) and age at follow up visits. Following \citet{proust2013analysis}, we consider the time covariate as the age minus $65$ years per $10$ years ($t = \frac{\text{age} - 65}{10}$). The MMSE test is skewed to left, hence we apply the $\log$ transformation and consider the marginals to be normal. For the BVRT test, we don't make any transformation and consider normal marginals as well. For the purpose of fitting, we re-scale the score of physical dependency (ordinal response) from $\{0,\dots,3\}$ to $\{1,\dots,4\}$. Here also for the two continuous responses we consider GLM model of the form -
\begin{equation}\label{creal2}
g(E(Y_{ij})) = \beta_0 + \text{sex}_i\beta_1 + \text{dem}_i\beta_2 + \text{cep}_i\beta_3 + t_{ij}\beta_4, \quad j = 1,\dots,n_i,
\end{equation}
where observed $y_{ij}$ is the continuous response at the $j$-th time for subject $i$. For the ordinal response we consider the latent variable model as
\begin{align}\label{oreal2}
Y_{ij} & = k \;\; \text{if} \;\; \gamma(k-1) \leq Z_{ij} <\gamma(k), \;\; k = 1,\dots,4, \nonumber \\ Z_{ij} & = \text{sex}_i\beta_1 + \text{dem}_i\beta_2 + \text{cep}_i\beta_3 + t_{ij}\beta_4 + \epsilon_{ij}, \quad j = 1,\dots,n_i,
\end{align}
where $\epsilon_{ij} (i.i.d) \sim N(0,1)$. Again, specifications of the factor copula models are similar to Section \ref{sec6}. While fitting these models, we also compare them with the corresponding random effect models. Our aim of this analysis is to describe the decline with age of the global cognitive ability measured by these psychometric tests and to evaluate the association within the longitudinal responses.
\begin{table}[]
    \centering
    \begin{small}
    \scalebox{1.0}{
    \tabcolsep = 0.18cm
    \begin{tabular}{|c|c c|c|c c|c|c c|}
    \hline
    \multicolumn{3}{|c}{$\log$ - MMSE (normal)} & \multicolumn{3}{|c}{BVRT (normal)} & \multicolumn{3}{|c|}{HIER (ordinal)} \\
    \hline
    \textbf{Parameters} & Est. & SE & \textbf{Parameters} & Est. & SE & \textbf{Parameters} & Est. & SE \\
    \hline
    $\beta_0$ & 3.4208 & 0.0330 & $\beta_0$ & 10.9296 & 0.1992 & - & - & - \\
    $\beta_1$ & 0.0047 & 0.0173 & $\beta_1$ & 0.2297 & 0.1536 & $\beta_1$ & -0.3077 & 0.0860 \\
    $\beta_2$ & -0.1758 & 0.0225 & $\beta_2$ & -0.6483 & 0.1493 & $\beta_2$ & 0.3503 & 0.0848 \\
    $\beta_3$ & 0.0717 & 0.0214 & $\beta_3$ & 1.4134 & 0.1599 & $\beta_3$ & -0.1579 & 0.0878 \\
    $\beta_4$ & -0.1240 & 0.0157 & $\beta_4$ & -0.6678 & 0.0821 & $\beta_4$ & 0.9184 & 0.0562 \\
    $\phi$ & 0.3209 & 0.0295 & $\phi$ & 2.1432 & 0.0427 & - & - & - \\
    - & - & - & - & - & - & $\gamma_1$ & 0.3094 & 0.1232 \\
    - & - & - & - & - & - & $\gamma_2$ & 1.6401 & 0.1338 \\
    - & - & - & - & - & - & $\gamma_3$ & 2.9992 & 0.1608 \\
    \hline
    \end{tabular}}
    \caption{Estimated marginal parameters and their standard errors of $3$ considered tests of the PAQUID data using the regression models in (\ref{creal2}) and (\ref{oreal2}) respectively.}
    \label{tab:paquidfit1}
    \end{small}
\end{table}
\begin{table}[]
    \centering
    \begin{small}
    \scalebox{1.0}{
    \tabcolsep = 0.18cm
    \begin{tabular}{|c|c|c|c|c|c|c|c|}
    \hline
    & \textbf{Copula} & \textbf{Parameters} & Est. & SE & Log-likelihood & AIC & BIC \\
    \hline
    & Gaussian & $\rho_1$ & 0.7366 & 0.0339 & -455.97 & 925.93 & 955.44 \\
    $\log$ - & $1$-factor & & & & & & \\
    MMSE & Student-$t$ & $\rho_1$ & 0.8488 & 0.0092 & -191.11 & 398.21 & 431.93 \\
    & $1$-factor ($\nu = 3$) & & & & & & \\
    & Gaussian & $\rho_1$ & 0.6228 & 0.0719 & -435.61 & 887.23 & 920.94 \\
    & $2$-factor & $\rho_2$ & 0.5821 & 0.0648 & & & \\
    & Student-$t$ & $\rho_1$ & 0.7725 & 0.0472 & \textbf{-169.47} & \textbf{356.94} & \textbf{394.87} \\ 
    & $2$-factor ($\nu = 3$) & $\rho_2$ & 0.7726 & 0.0426 & & & \\
    \hline
    & Gaussian & $\rho_1$ & 0.5460 & 0.0249 & -4774.56 & 9563.11 & 9592.61 \\
    BVRT & $1$-factor & & & & & & \\
    & Student-$t$ & $\rho_1$ & 0.5678 & 0.0232 & \textbf{-4766.38} & \textbf{9548.76} & \textbf{9582.47} \\
    & $1$-factor ($\nu = 6$) & & & & & & \\
    & Gaussian & $\rho_1$ & 0.4036 & 0.0096 & -4774.65 & 9565.30 & 9599.02 \\
    & $2$-factor & $\rho_2$ & 0.4011 & 0.0376 & & & \\
    & Student-$t$ & $\rho_1$ & 0.4209 & 0.0662 & -4766.33 & 9550.70 & 9588.63 \\
    & $2$-factor ($\nu = 8$) & $\rho_2$ & 0.4199 & 0.0696 & & & \\
    \hline
    & Gaussian & $\rho_1$ & 0.7282 & 0.0183 & -2154.38 & 4324.76 & 4358.48 \\
    HIER & $1$-factor & & & & & & \\
    & Student-$t$ & $\rho_1$ & 0.7331 & 0.0185 & -2139.54 & 4297.07 & 4335.00 \\
    & $1$-factor ($\nu = 4$) & & & & & & \\
    & Gaussian & $\rho_1$ & 0.5249 & 0.0146 & -2149.77 & 4317.54 & 4355.48 \\
    & $2$-factor & $\rho_2$ & 0.6192 & 0.0218 & & & \\
    & Student-$t$ & $\rho_1$ & 0.6083 & 0.0426 & \textbf{-2124.43} & \textbf{4268.85} & \textbf{4311.00} \\
    & $2$-factor ($\nu = 3$) & $\rho_2$ & 0.5611 & 0.0680 & & & \\
    \hline
    \end{tabular}}
    \caption{Estimated dependence parameters and their standard errors of $3$ considered tests of the PAQUID data with $1$-factor and $2$-factor copula models. Maximum log-likelihood value, AIC and BIC for each model are reported.}
    \label{tab:paquidfit2}
    \end{small}
\end{table}
\begin{table}[]
    \centering
    \begin{small}
    \scalebox{1.0}{
    \tabcolsep = 0.18cm
    \begin{tabular}{|c|c c|c|c c|c|c c|}
    \hline
    \multicolumn{3}{|c}{$\log$ - MMSE (normal)} & \multicolumn{3}{|c}{BVRT (normal)} & \multicolumn{3}{|c|}{HIER (ordinal)} \\
    \hline
    \textbf{Parameters} & Est. & SE & \textbf{Parameters} & Est. & SE & \textbf{Parameters} & Est. & SE \\
    \hline
    $\beta_0$ & 3.4308 & 0.0257 & $\beta_0$ & 10.7646 & 0.1868 & - & - & - \\
    $\beta_1$ & 0.0065 & 0.0185 & $\beta_1$ & 0.1048 & 0.1452 & $\beta_1$ & -0.2736 & 0.0860 \\
    $\beta_2$ & -0.1804 & 0.0195 & $\beta_2$ & -0.5950 & 0.1561 & $\beta_2$ & 0.2572 & 0.0933 \\
    $\beta_3$ & 0.0670 & 0.0202 & $\beta_3$ & 1.4516 & 0.1577 & $\beta_3$ & -0.1860 & 0.0787 \\
    $\beta_4$ & -0.1346 & 0.0104 & $\beta_4$ & -0.6690 & 0.0679 & $\beta_4$ & 1.0603 & 0.0361 \\
    $V[b]$ & 0.0157 & 0.0024 & $V[b]$ & 0.8749 & 0.1597 & $V[b]$ & 0.5597 & 0.0625 \\
    $\phi$ & 0.2963 & 0.0049 & $\phi$ & 1.8019 & 0.0306 & - & - & - \\
    - & - & - & - & - & - & $\gamma_1$ & 0.3091 & 0.1242 \\
    - & - & - & - & - & - & $\gamma_2$ & 1.5805 & 0.1422 \\
    - & - & - & - & - & - & $\gamma_3$ & 3.0121 & 0.1577 \\
    \hline
    \multicolumn{3}{|c}{Log-likelihood \quad \textbf{-593.76}} & \multicolumn{3}{|c}{Log-likelihood \quad \textbf{-4777.94}} & \multicolumn{3}{|c|}{Log-likelihood \quad \textbf{-2251.08}} \\
    \multicolumn{3}{|c}{AIC \quad \textbf{1201.52}} & \multicolumn{3}{|c}{AIC \quad \textbf{9569.87}} & \multicolumn{3}{|c|}{AIC \quad \textbf{4512.17}} \\
    \multicolumn{3}{|c}{BIC \quad \textbf{1231.02}} & \multicolumn{3}{|c}{BIC \quad \textbf{9609.89}} & \multicolumn{3}{|c|}{BIC \quad \textbf{4533.24}} \\
    \hline
    \end{tabular}}
    \caption{Estimated parameters and their standard errors of $3$ considered tests of the PAQUID data by adding random intercepts to the regression models in (\ref{creal2}) and (\ref{oreal2}) respectively. Maximum log-likelihood value, AIC and BIC for each model are reported.}
    \label{tab:paquidfit3}
    \end{small}
\end{table}
\par
\begin{figure}
    \centering
    \includegraphics[width=15cm]{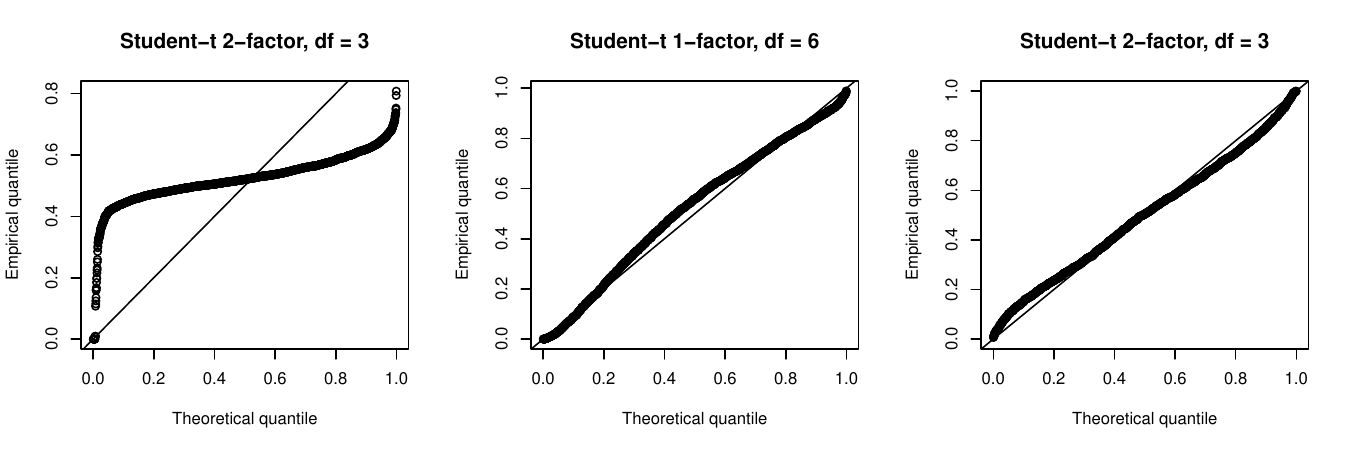}
    \caption{Uniform probability plots of the residuals of the best fitting copula models for (i) MMSE, (ii) BVRT psychometric tests and (iii) HIER for PAQUID data set.}
    \label{fig:resid.paq1}
\end{figure}

 Table \ref{tab:paquidfit1} and \ref{tab:paquidfit2} presents the marginal parameter estimates and the dependence parameter estimates of the PAQUID data under different factor copula models, respectively. Based on the marginal estimates, gender is not significant for MMSE and BVRT tests. As previously depicted in \ref{fig:profile.paq1} and from the estimates of $\beta_2$ in all the models, subjects initiated with positive diagnosis of dementia, subsequently performed poorer in $2$ two of the psychometric tests and their score of physical dependency (HIER) increased more compared to others. Based estimates of $\beta_3$, in all the models, subjects who had higher educational level had higher cognitive ability and lower physical dependency, compared to others through out the cohort. As seen by the estimates of $\beta_4$, increasing age had significant impact on psychometric tests. Student-$t$ $2$-factor copula models provide with the best fit for the MMSE test and the score of physical dependency. For the BVRT test $1$-factor Student-$t$ copula model best describes the underlying dependence. The correlation between the latent variables with each responses is strong and positive as well. The parameter estimates from the considered random effect models are given in Table \ref{tab:paquidfit3}. The estimates of the fixed effects are quite similar with the marginal models. We see that factor copula models outperforms the random effect models for each tests in this data set based on the selection criteria. This implies considered factor copula models better explains the temporal dependency in this data set. Figure \ref{fig:resid.paq1} presents the residual plots of uniform scale for each psychometric tests corresponding to their best fitting model. The models for BVRT test and the score of physical dependency suggests perfect fitting models but for the MMSE test it shows substantial lack of fit even though Student-$t$ 2-factor copula provided with the best fit among other competitors.
Mixed models can be affected by the misspecification of random effects distribution (\citet{litiere2007type}). But on the other hand factor copula models have more direct interpretation of the underlying dependence mechanism. Of course one can choose between random effects and copula models based on the scientific question of interest what the investigator is looking for.
\section{Discussion}\label{sec8}
In this article, we introduced factor copula models tailored for arbitrary non-Gaussian longitudinal data, incorporating covariates. Our proposed models have demonstrated efficacy in modeling unbalanced longitudinal data featuring both discrete and continuous responses. By adopting parsimonious specifications for factor copula parameters, our models exhibit scalability, allowing for seamless accommodation of dependence in moderate to high dimensions without encountering computational hurdles. We employed the two-stage Inference Function for Margins (IFM) method to estimate the model parameters. Through simulation studies, we observed consistent and reliable estimation of both the marginal and dependence parameters of the models. Furthermore, we conducted comparisons between our proposed models and widely used random effect models, employing similar specifications for fixed covariates. There are further scopes of extending this endevour to other bivariate copulas as well.
\par Factor copula models inherently assume a homogeneous dependence structure for all subjects, which may not always align with reality, especially in scenarios where measurements taken closer in time exhibit stronger dependence compared to those taken further apart. Thus, there is room for further refinement of factor copula models to accommodate potential time-heterogeneity. For illustrative purposes, we examined two widely used real-world datasets prevalent in joint modeling literature. A pertinent statistical inquiry involves evaluating the contemporaneous association between each longitudinal response. This can be approached in two ways. Firstly, one can employ another multivariate copula (such as Gaussian or Student-$t$) to construct the joint distribution of responses at each time point. This methodology has been employed by \citet{baghfalaki2020transition}, who utilized a Gaussian copula alongside a transition model to describe temporal dependence. Similarly, \citet{sefidi2022analysis} adopted a similar strategy, employing a bivariate copula to associate continuous and ordinal responses, while leveraging D-vine copulas to account for temporal dependence. Building upon these methodologies, one can incorporate multivariate copulas to associate the subject-specific latent variables in factor copulas, thereby incorporating correlated latent variables on each factor. Consequently, all dependence and association parameters can be estimated through the second stage of IFM estimation using the joint likelihood. Alternatively, a more sophisticated approach involves incorporating correlated random effects in the marginals of each longitudinal response to explain contemporaneous associations between responses. This avenue represents a focal point for our future investigations within this framework. \\
\section{Declarations}
\textbf{Conflict of interest}: The author declares no Conflict of interest.
\bibliographystyle{agsm}
\bibliography{paper4ref}
\end{document}